\documentclass[reprint,amsmath,amssymb,aps,superscriptaddress]{revtex4-2}

\usepackage{dcolumn}
\usepackage{bm}
\usepackage{graphicx} 
\usepackage{amsmath}
\usepackage[version=4]{mhchem} 
\usepackage{siunitx}
\usepackage{miller}

\usepackage[colorlinks,plainpages=false,linkcolor=blue,urlcolor=blue,citecolor=blue,pdfpagemode=UseNone,pdfstartview=FitBH]{hyperref} 
\usepackage[Charter,greekuppercase=italicized]{mathdesign} 

\newcommand{\sn}{\ce{SnTaSe2}}
\newcommand{\pb}{\ce{PbTaSe2}}
\newcommand{\pbsn}{\ce{Pb_{1-x}Sn_{x}TaSe2}}

\begin{document}


\title{Enhancement of the superconducting transition temperature due to multiband effect in the topological nodal-line semimetal \ce{Pb_{1-x}Sn_{x}TaSe2}}

\author{K. Kumarasinghe}
\author{A. Rahman}
\author{M. Tomlinson}
\author{D. Le}
\affiliation{Department of Physics, University of Central Florida, Orlando, Florida 32816, USA}
\author{F. Joshua}
\author{L. Zhai}
\affiliation{NanoScience Technology Center, University of Central Florida, Orlando, Florida 32826, USA}
\affiliation{Department of Chemistry, University of Central Florida, Orlando, Florida 32816, USA}
\author{Y. Nakajima}
\email[Corresponding author: ]{Yasuyuki.Nakajima@ucf.edu}
\affiliation{Department of Physics, University of Central Florida, Orlando, Florida 32816, USA}

\begin{abstract}

We report a systematic study of the normal-state and superconducting properties of single crystal \pbsn\ $(0\leq x \leq 0.23)$. Sn doping enhances the superconducting temperature $T_{c}$ up to 5.1 K while also significantly increasing impurity scattering in the crystals. For $x=0$ and 0.018, the specific heat jump at $T_{c}$ exceeds the Bardeen-Cooper-Schrieffer (BCS) weak-coupling value of 1.43, indicating the realization of strong-coupling superconductivity in undoped and slightly Sn-doped \pb. Substituting Pb with more Sn lowers the specific heat jump at $T_{c}$ below the BCS value of 1.43, which cannot be explained by a single-gap model. Rather, the observed specific heat data of moderately Sn-doped \pb\ ($x= 0.08$ and 0.15) are reproduced by a two-gap model. Our density functional theory calculations suggest that three-dimensional Fermi pockets appear due to a reduction of the spin-orbit gap with Sn doping, and the multiband effect arising from these emergent Fermi pockets enhances the effective electron-phonon coupling strength, leading to the increase in $T_{c}$ of \pbsn.

\end{abstract}

\maketitle

\section{Introduction}

Noncentrosymmetric superconductors have attracted great interest because of their exotic superconducting properties, including an upper critical field that exceeds the Pauli paramagnetic limit \cite{bauer04,kimur07} and the realization of unconventional superconductivity \cite{yuan06,nishi07,chen11}. The intrinsic lack of the inversion symmetry results in the antisymmetric spin-orbit coupling (SOC) and allows the mixture of spin-singlet (even-parity) and spin-triplet (odd-parity) pairing in the superconducting gap function, even in the conventional electron-phonon coupling mechanism. The mixture of the parities can be finely tuned by adjusting the strength of the antisymmetric SOC. Notably, the noncentrosymmetric superconductor \ce{Li2(Pd_{1-x}Pt_x)3B} exhibits a striking crossover from an even-parity-dominant fully-gapped superconductivity to an odd-parity-dominant nodal superconductivity by controlling Pt concentration \cite{harad12}. Moreover, noncentrosymmetric superconductors present a promising avenue for realizing topological superconductivity, which can host Majorana-bound states \cite{sato17}.

The topological nodal-line semimetal \pb\ is one of the intriguing examples of noncentrosymmetric superconductors. Its unique crystal structure—characterized by alternating layers of Pb and \ce{TaSe2} along the $c$ axis—lacks inversion symmetry. This material shows superconductivity at the superconducting transition temperature $T_{c}$ =3.6--3.8 K \cite{zhang2016superconducting,wang15b,ali2014noncentrosymmetric}. Extensive experimental study, including specific heat \cite{zhang2016superconducting}, thermal conductivity \cite{wang16}, penetration depth \cite{pang16}, and $\mu$SR measurements \cite{wilson2017mu}, have revealed the fully-gapped superconducting state in \pb. Angular-resolved-photoemission spectroscopy has uncovered bulk nodal lines with nontrivial band topology and fully spin-polarized topological surface states, stemming from the asymmetric SOC and protected by reflection symmetry \cite{chang16,bian2016topological,wang15b}. The interplay between the bulk fully-gapped superconductivity and the topological surface state can stabilize topological superconductivity through their proximity effect \cite{fu08}. In addition, \sn, with a weaker SOC strength than \pb, is also predicted to have topological nodal lines and show superconductivity at $T_{c} = 5.7$ K \cite{chen16a}. Thus, exploring the evolution of superconducting properties by tuning the SOC strength through substituting Pb with Sn in \pb\ presents an exciting opportunity to realize an exotic superconducting state.

We here report our findings on the normal-state and superconducting properties of single crystal \pbsn. Our results show that Sn doping significantly enhances $T_{c}$ while introducing notable disorder into the crystals, suggesting the robustness of superconductivity against disorder. We observe a slight increase in the Debye temperature estimated from the resistivity and specific heat as Sn doping levels increase. The specific heat jumps at $T_{c}$ for $x=0$ and 0.018 exceed the Bardeen-Cooper-Schrieffer (BCS) weak-coupling value of 1.43, indicating the realization of strong-coupling superconductivity in undoped \pb\ and slightly Sn-doped \pb. On the other hand, more Sn doping suppresses the specific heat jump at $T_{c}$ less than the BCS value, which cannot easily be explained by the single-gap model. Instead, our specific heat data for moderately Sn-doped \pb\ agree with the two-gap model. Supported by density functional theory (DFT) calculations, our observations suggest that three-dimensional Fermi pockets arise from a reduced spin-orbit gap due to Sn substitution, leading to a multiband superconductivity that enhances $T_{c}$.

\section{METHODS}
\begin{figure}[t]
\includegraphics[width=0.48\textwidth]{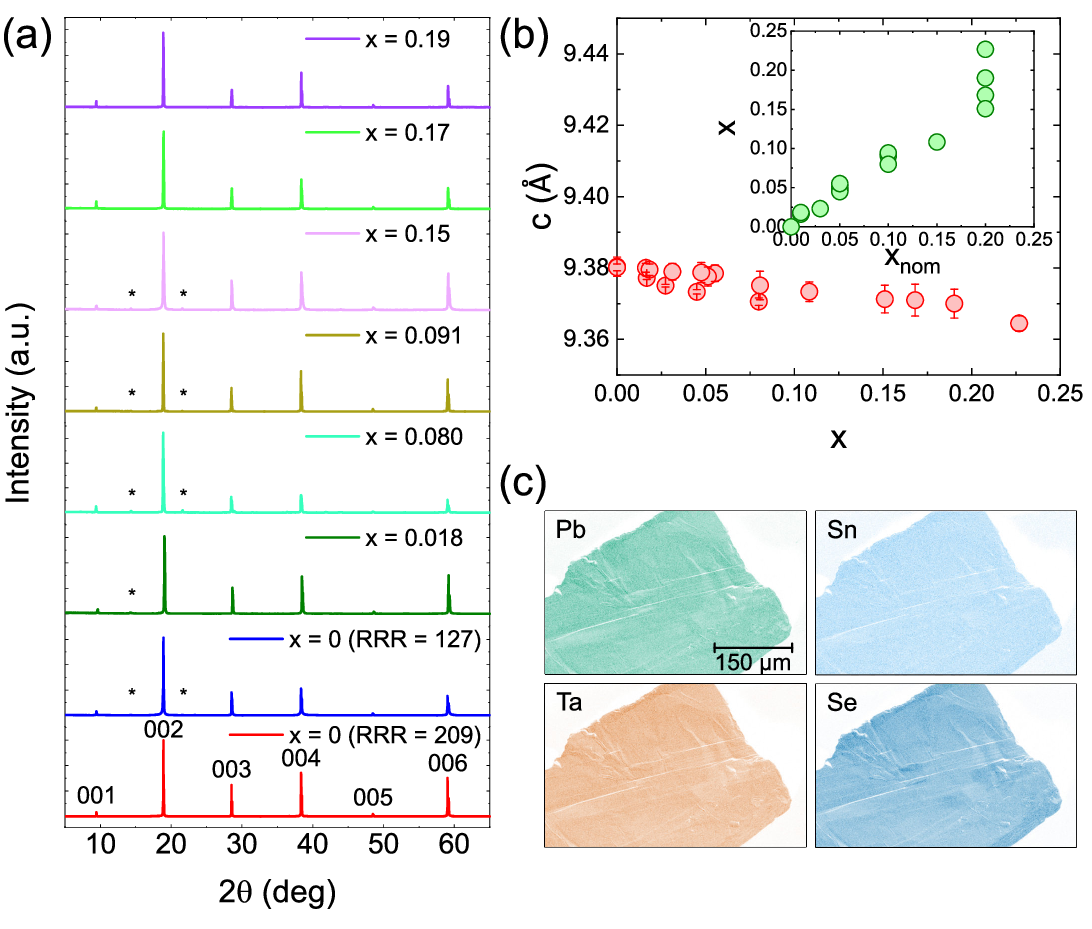}
\caption{\label{fig:xrd}(a) X-ray diffraction pattern for \ce{Pb_{1-x}Sn_{x}TaSe2} using Cu K$\alpha$ radiation. Asterisks indicate peaks for misfit material \ce{(PbSe)_{1.12}TaSe2}. (b) Lattice parameter $c$, determined by (00$\ell$) peaks in x-ray diffraction pattern, as a function of Sn concentration $x$. The Sn concentration $x$ is determined indivisually for each sample by XRF spectroscopy. Inset: actual Sn concentration $x$ vs nominal Sn concentration $x_{\mathrm{nom}}$. (c) EDS maps of Pb, Sn, Ta, and Se elements for $x=0.18$.}
\end{figure}

Single crystals of \ce{Pb_{1-x}Sn_{x}TaSe2} were grown using the chemical vapor transport method. Polycrystalline \ce{Pb_{1-x}Sn_{x}TaSe2} was first synthesized by the solid-state reaction. High-purity starting elements of Pb, Sn, Ta, and Se were placed in an alumina crucible with the molar ratio of Pb:Sn:Ta:Se = 1-$x_{\mathrm{nom}}$:$x_{\mathrm{nom}}$:1:2, where $x_{\mathrm{nom}}$ is a nominal Sn concentration, and sealed inside an evacuated quartz tube. The quartz ampoule was heated at 850 \si{\celsius} for 5 days to obtain polycrystals. The obtained polycrystals were ground and placed at one end of a quartz tube along with the transport agent \ce{PbCl2} of 3 mg/cm$^{3}$. The quartz ampoule was then evacuated, sealed, and placed in a two-zone furnace. The hot and cold ends of the furnace were maintained at 850 \si{\celsius} and 800 \si{\celsius}, respectively, for 3 weeks. Large, shiny, flake-like single crystals were obtained at the cold end. The typical size of undoped PbTaSe$_{2}$ is  $2.5\times 1.5\times 0.15$ mm$^{3}$ while those of Pb$_{1-x}$Sn$_{x}$TaSe$_{2}$ with higher doping are smaller, $2\times 0.5\times 0.05$ mm$^{3}$. No discernible change in color with doping was observed. The actual Sn concentration $x$ for each sample was determined individually using x-ray fluorescence (XRF) spectroscopy. The relation between the nominal and actual Sn concentrations is plotted in the inset to FIG.\ref{fig:xrd}(b). The spatial distribution of elements was investigated using a scanning electron microscope (SEM) equipped with energy-dispersive x-ray spectroscopy (EDS). The crystal structures of the obtained single crystals were characterized using x-ray diffraction.

We performed the resistivity measurements with an AC resistance bridge using the standard four-wire configuration. Currents of 100 - 316 $\mu$A were applied along the $ab$ plane. Specific heat measurements were conducted using the long-relaxation method \cite{taylo07} with a homemade calorimeter.

DFT calculations are performed using the Vienna \emph{Ab initio} Simulation Package (VASP 5.4.4) \cite{RN429, RN425}, employing the projector-augmented wave (PAW) pseudopotential method \cite{RN424, RN449} and a plane wave basis set. We use the generalized-gradient approximation (GGA) in the form of Perdew-Berke-Enzerhoff (PBE) exchange-correlation functionals \cite{RN453, RN435} for accounting for the exchange correlation. We set a cutoff energy of 500 eV for plane-wave expansion. All electronic iterations are converged with 0.001 meV threshold. The internal coordinates of atoms and lattice constant of the bulk structures are optimized so that forces acting on each atom are less than 5 meV/\AA\, and the stress is minimized to less than 0.2 kbar. We use the Gaussian smearing method with $\sigma$ = 0.1 eV and sample the Brillouin Zone with a 18$\times$18$\times$6 $\Gamma$-centered grid for \ce{SnTaSe2} and \ce{PbTaSe2} and with a 9$\times$9$\times$6 $\Gamma$-centered grid for \ce{Pb_{0.75}Sn_{0.25}TaSe2}. \ce{Pb_{0.75}Sn_{0.25}TaSe2} is modeled with 2$\times$2$\times$1 supercell of \ce{PbTaSe2} with one Pb atom that is replaced by Sn atom. Spin-orbit coupling is incorporated for the calculations of electronic band structures of the systems. The Fermi surfaces of \ce{SnTaSe2} and \ce{PbTaSe2} are calculated with the IFermi package \cite{RN23109}. The electronic band structure of \ce{Pb_{0.75}Sn_{0.25}TaSe2} was unfolded using the VaspBandUnfolding package \cite{RN23110}.

\begin{figure}[t]
\includegraphics[width=0.48\textwidth]{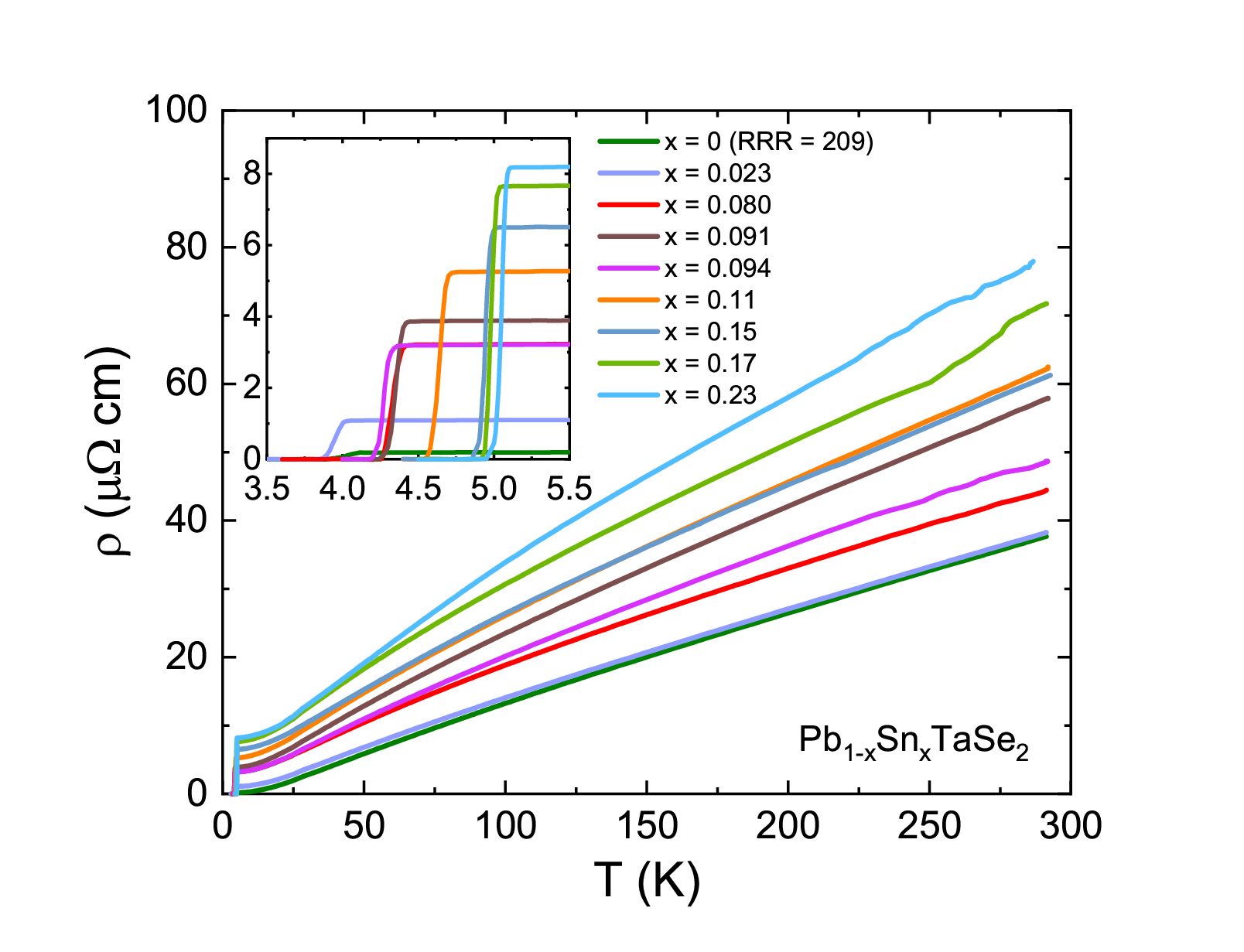}
\caption{\label{fig:rho}Overall temperature dependence of the in-plane resistivity of \ce{Pb_{1-x}Sn_{x}TaSe2} at $\mu_{0}H$ = 0 T. Both undoped and Sn-doped \pb\ show metallic behavior and undergo a superconducting transition at low temperatures. The inset shows low-temperature resistivity of \pbsn.}
\end{figure}

\begin{figure}[t]
\includegraphics[width=0.35\textwidth]{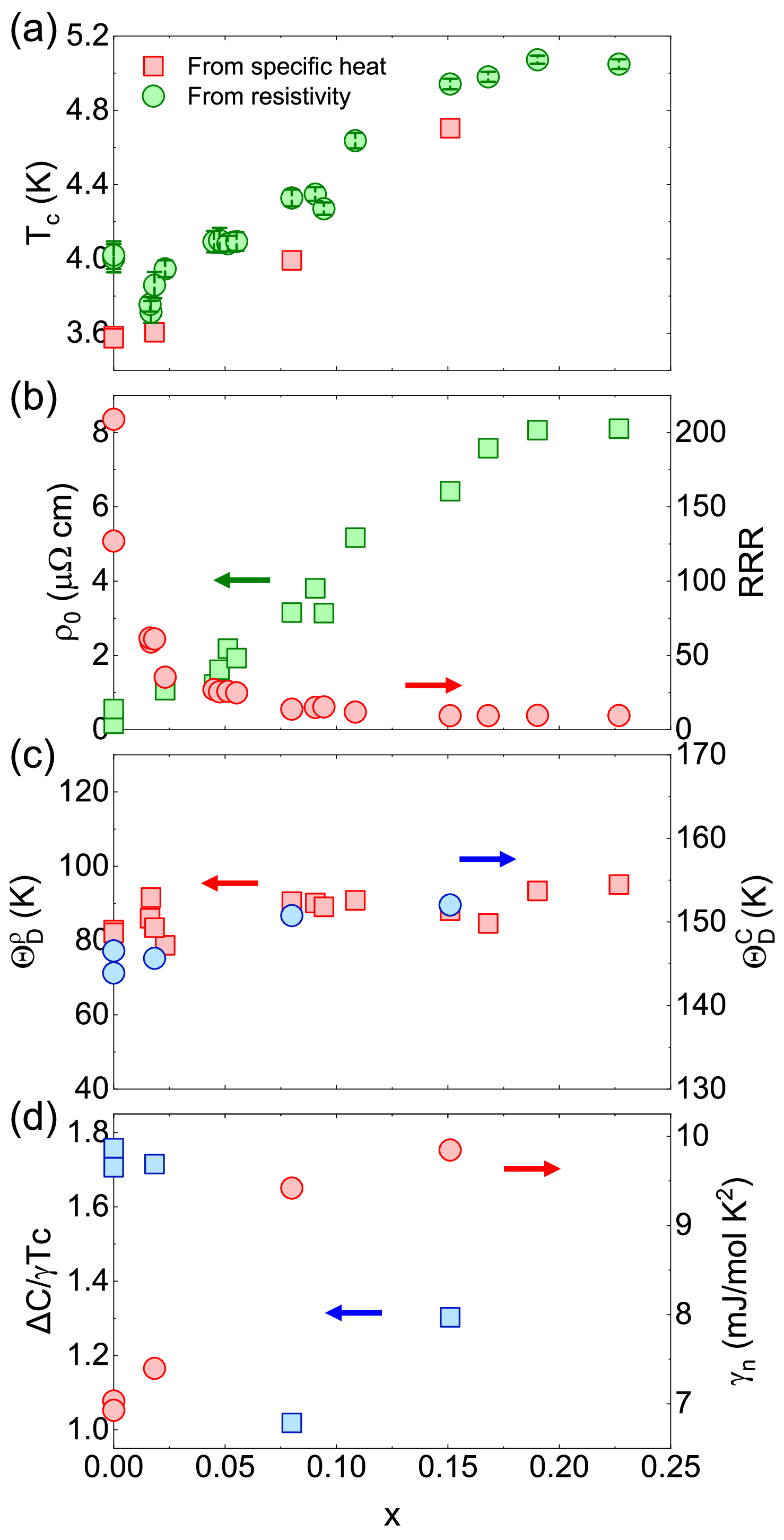}
\caption{\label{fig:tc} (a) Superconducting transition temperatures $T_{c}$ of \pbsn\ obtained from the resistivity (green circles) and the specific heat (red squares) as a function of Sn concentration $x$. $T_{c}$ is defined by the midpoint of resistive transitions and by the local entropy balance in the specific heat measurements. The error bars of $T_{c}$ obtained from the resistivity represent 10 \% and 90 \% criteria. (b) The residual resistivity ratio RRR = $\rho(290 \mathrm{K})/\rho(T^{\mathrm{onset}}_{c})$ (red circles) and the residual resistivity $\rho_{0}$ (green squares) versus $x$. $\rho_{0}$ is extracted from a fit to the data below 7 K, using $\rho(T) = \rho_{0} + AT^{2}$. (c) Debye temperature $\Theta_{\mathrm{D}}^{C}$ obtained from the specific heat (blue circles) and $\Theta_{\mathrm{D}}^{\rho}$ obtained from the resistivity (red squares) versus Sn concentration $x$. $\Theta_{\mathrm{D}}^{C}$ was estimated using $\Theta_{\mathrm{D}}=(12\pi^{4}nR/5\beta_{\mathrm{ph}})^{1/3}$ and $\Theta_{\mathrm{D}}^{\rho}$ is extracted from a fit to the resistivity data using the Bloch--Gr{\"u}neisen formula. (d) Normalized specific heat jump $\Delta C/\gamma_{\mathrm{n}}T_{c}$ at $T_{c}$ (blue squares) and the normal-state Sommerfeld coefficient $\gamma_{\mathrm{n}}$ (red circles) as a function of $x$. $\gamma_{\mathrm{n}}$ is extracted from a fit to the data in an applied magnetic field of $\mu_{0}H$ = 1 T, using $C_{\mathrm{n}}/T = \gamma_{\mathrm{n}} +\beta_{\mathrm{ph}} T^{2} + \delta_{\mathrm{ph}} T^{4}$.}
\end{figure}

\section{Results and Discussion}

We observe x-ray diffraction patterns consistent with the noncentrosymmetric crystal structure of \ce{PbTaSe2} with the space group $\textit{P}\bar{6}m2$ for the samples in the whole doping range, as shown in FIG.\ref{fig:xrd}(a). Tiny peaks around $2\theta =\ang{14}$ and \ang{21.5} indicated by asterisks are attributed to stacking faults that form the PbSe layers, present in the misfit material \ce{(PbSe)_{1.12}TaSe_{2}} ($T_{c}=1.3$ K) \cite{yang18c}. As discussed later, the presence of a minor impurity phase does not affect the bulk superconductivity. The lattice constant $c$ obtained from (00$\ell$) peaks as a function of $x$ is plotted in FIG.\ref{fig:xrd}(b). As expected for the substitution of Sn for Pb, $c$ decreases with $x$.  

To examine the spatial distribution of elements, we conducted SEM-EDS analysis. As shown in FIG.\ref{fig:xrd}(c), the EDS map of Pb, Sn, Ta, and Se elements for x = 0.18 ($x_{\mathrm{nom}}=0.2$) shows no discernible  macroscopic inhomogeneity of Sn doping within a sample.

Sn doping enhances $T_c$ of \pb. We present the in-plane resistivity $\rho$ of \pbsn\ single crystals as a function of temperature ($T$) in zero magnetic field (FIG.\ref{fig:rho}). For all  Sn concentrations $x$, the resistivity exhibits metallic behavior as the temperature decreases.  At high temperatures, the resistivity shows linear-in-$T$ behavior, indicating the dominant electron-phonon scattering.

Regardless of the Sn concentrations, we observe sharp superconducting transitions at low temperatures, as shown in the inset to FIG.\ref{fig:rho}. $T_{c}$ were determined by the midpoint of the resistive transition. For \pb, superconductivity occurs at $T_{c}$ = 4.02 K, which is slightly higher than the previously reported values of 3.6-3.8 K from the resistivity measurements \cite{sankar2017anisotropic,bian2016topological,zhang2016superconducting}. We plot $T_{c}$ obtained from the resistivity measurements as a function of the Sn doping concentration $x$ in FIG.\ref{fig:tc} (a) (green circles). Notably, $T_{c}$ increases with Sn doping after showing a slight dip at around $x=0.02$. We will discuss the origin of this dip later. Above $x=0.19$, $T_{c}$ reaches $\sim 5.1$ K, by $\sim$1.1 K or 26 \% higher than the value for $x=0$.

Sn doping induces disorder in \pbsn\ while simultaneously increasing $T_{c}$. To extract the residual resistivity $\rho_{0}$, we fit the low-temperature resistivity from $T_{c}$ to 7 K, using $\rho(T) = \rho_{0} + AT^{2}$, where $\rho_{0}$ is the residual resistivity and $A$ is a coefficient of electron-electron scattering term. As shown in FIG.\ref{fig:tc} (b) (green squares), $\rho_{0}$ increases from 0.16 $\mu\Omega$ cm to 8 $\mu\Omega$ cm, a factor of 50, with increasing Sn concentration $x$. To eliminate uncertainties stemming from geometric factors in calculating the absolute values of the resistivity, we also evaluate the residual resistivity ratio RRR as a scattering measure. The RRR is defined as RRR $=\rho(290~\mathrm{K})/\rho(T^{\mathrm{onset}}_{c})$ and is plotted as a function of $x$ in FIG.\ref{fig:tc}(b) (red circles). The RRR shows a sharp drop at a slight doping of 2\% Sn into \pb\ and gradually decreases with increasing Sn concentration. The undoped \pb\ has the largest RRR of 209, while the sample with $x=0.23$ has the smallest RRR of 9.5, approximately 20 times smaller than that of the undoped sample. Both $\rho_{0}$ and RRR indicate that the Sn doping introduces disorders in the system.

To investigate the effect of Sn doping on the electron-phonon coupling, we evaluate the Debye temperature $\Theta_{\mathrm{D}}$. The temperature dependence of the normal-state resistivity due to the electron-phonon scattering can be described by the Bloch--Gr{\"u}neisen formula, 
\begin{equation}
\rho(T) = \rho_{0}+A_{\mathrm{BG}}\left(\frac{T}{\Theta_{\mathrm{D}}}\right)^5\int_{0}^{\frac{\Theta_{\mathrm{D}}}{T}}\frac{x^{5}dx}{(e^{x}-1)(1-e^{-x})},\label{eq:BG}
\end{equation}
where $A_{\mathrm{BG}}$ is a constant related to the electron-phonon coupling strength \cite{ziman01}. $\Theta_{\mathrm{D}}$, obtained from fits to the resistivity data in the entire temperature range using Eq.(\ref{eq:BG}), is plotted in FIG.\ref{fig:tc}(c) (red squares). For $x=0$, $\Theta_{\mathrm{D}}$ is $\sim$ 83 K, consistent with previously reported values \cite{sankar2017anisotropic,long2016single}. $\Theta_{\mathrm{D}}$ shows weak dependence on $x$ and exhibits a slight increase by 6\% at $x=0.23$, which is qualitatively consistent with substituting heavier Pb with the lighter Sn.

\begin{figure}[t]
\includegraphics[width=0.48\textwidth]{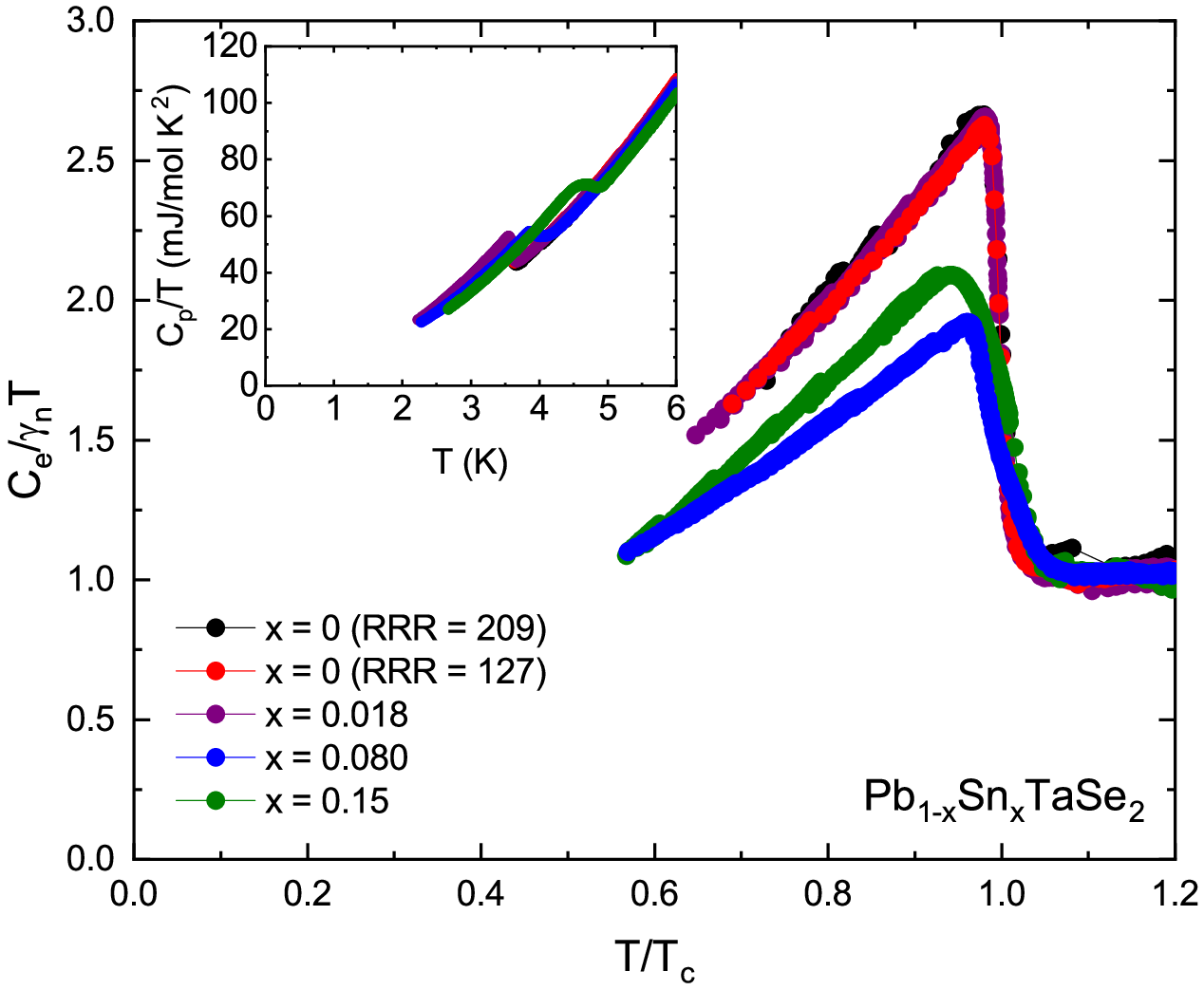}
\caption{\label{fig:ce} Normalized electronic specific heat $C_{\mathrm{e}}(T)/\gamma_{\mathrm{n}}T$ of \pbsn\ as a function of the reduced temperature $T/T_{c}$. Inset: temperature dependence of $C_{p}/T$ of \pbsn.}
\end{figure}

We confirm bulk superconductivity in \pbsn\ through specific heat measurements. The temperature dependence of the specific heat $C_{p}(T)$ at 0 T for \pbsn\ is shown in the inset to FIG.\ref{fig:ce}. In the normal state at high temperatures, the $C_{p}$ of \pbsn\ overlap, suggesting that Sn doping has minimal effect on the lattice contribution to the specific heat. Upon lowering temperatures, we observe sharp jumps associated with the superconducting transitions at $T_{c}$. $T_{c}$ was determined using the local entropy balance at the superconducting transition. $T_{c}$ obtained from the specific heat is plotted as a function of $x$ in FIG.\ref{fig:tc}(a) (red squares). $T_{c}$ of undoped \pb\ with RRR = 209 and RRR = 127 are 3.58 K and 3.57 K, respectively. These values are in good agreement with previously reported values obtained from thermodynamic measurements, including specific heat \cite{zhang2016superconducting} and magnetic susceptibility measurements \cite{wilson2017mu}. However, for undoped \pb, $T_{c}$ obtained from the specific heat deviates from those obtained from the resistive transitions. In contrast, at the higher Sn concentrations, $T_{c}$ obtained from the specific heat are comparable to those obtained from the resistivity measurements. We attribute this discrepancy between $T_{c}$ determined by resistivity and specific heat for undoped {\pb} to an extrinsic effect---surface superconductivity. While specific heat detects bulk superconductivity, resistivity can be susceptible to surface superconductivity. Notably, despite the extremely high RRR > 200, the resistive transition width of the undoped sample is broader compared to Sn-doped samples with lower RRR. This broadening likely results from superconducting fluctuations \cite{baeva24}, which are amplified by the reduced dimensionality of surface superconductivity. Therefore, the dip in $T_{c}$ around $x = 0.02$ determined by resistivity is extrinsic rather than due to the impurity effect on $T_{c}$, as observed in unconventional superconductors.

\begin{figure}[htb]
\includegraphics[width=0.33\textwidth]{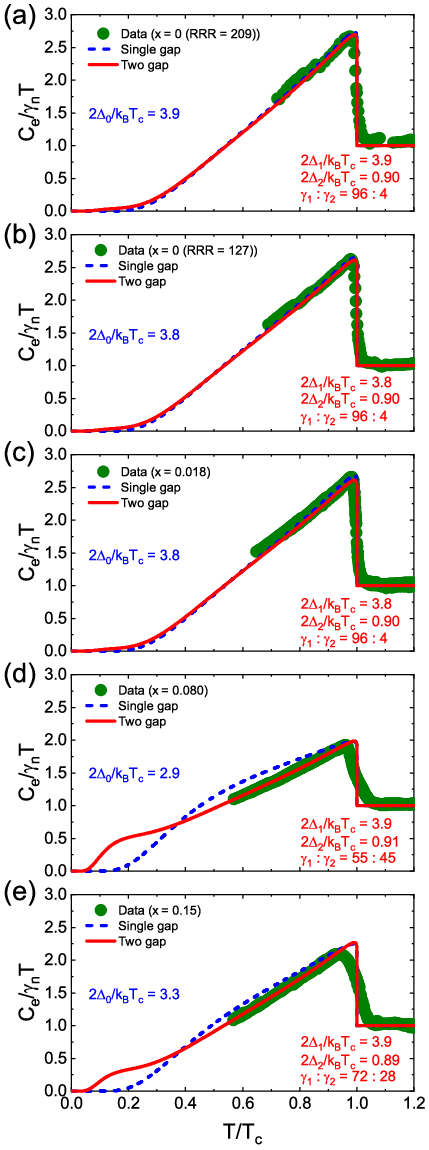}
\caption{\label{fig:gap} $C_{\mathrm{e}}(T)/\gamma_{\mathrm{n}}T$ vs $T/T_{c}$ for (a) $x=0$ (RRR=209), (b) $x=0$ (RRR=127), (c) $x=0.018$, (d) $x=0.080$, and (e) $x=0.15$. The red solid curves are the fits to each data set using the two-gap model. For $x=0$ and 0.018, the blue dashed lines represent the best fit to the data using the single-gap model. For $x=0.08$ and 0.15, the blue dashed lines are theoretical curves based on the single-gap model, only reproducing the specific heat jump size at $T_{c}$.}
\end{figure}

We take a closer look at the effect of Sn doping on the normal-state specific heat of \pbsn. To extract the electronic and lattice contributions to the specific heat, we fit the normal-state specific heat data using the following equation:
\begin{equation}
C_{\mathrm{n}}(T)/T = \gamma_{\mathrm{n}} + C_{\mathrm{ph}}(T)/T,
\end{equation}
where $\gamma_{\mathrm{n}}$ is the Sommerfeld coefficient and $C_{\mathrm{ph}}(T) = \beta_{\mathrm{ph}} T^{3} + \delta_{\mathrm{ph}} T^{5}$ represents the phonon contribution to the specific heat, which is assumed to be independent of applied magnetic field. We obtain the normal-state specific heat $C_{\mathrm{n}}(T)$ by applying a magnetic field of 1 T along the $c$ axis, which completely suppresses superconductivity in the measured temperature range for all \pbsn. The parameters obtained for \pb\ with RRR = 127 are $\gamma_{\mathrm{n}}$ = 6.93 mJ/mol K$^2$, $\beta_{\mathrm{ph}}$ = 2.61 mJ/mol K$^4$, and $\delta_{\mathrm{ph}}$ = 7.01$\times 10^{-3}$ mJ/mol K$^6$, which are in good agreement with previously reported values \cite{sankar2017anisotropic,zhang2016superconducting,ali2014noncentrosymmetric,long2016single,sun2020fully}.
As shown in FIG.\ref{fig:tc}(d) (red circles), the obtained $\gamma_{\mathrm{n}}$ increases with increasing Sn concentration.

The Debye temperature $\Theta_{\mathrm{D}}$ is estimated from the extracted $\beta_{\mathrm{ph}}$ using $\Theta_{\mathrm{D}}=(12\pi^{4}nR/5\beta_{\mathrm{ph}})^{1/3}$, where n = 4 is the number of atoms per formula unit and $R = 8.314$ J/mol K is the molar gas constant. As plotted in FIG.\ref{fig:tc}(c) (blue circles), the obtained $\Theta_{\mathrm{D}}$ is $\sim$ 145 K for $x=0$ and weakly dependent on $x$, enhanced only by 3\% at $x=0.15$. $\Theta_{\mathrm{D}}$ obtained from the specific heat measurements is about twice as large as that obtained from the resistivity measurements. This discrepancy arises from the different temperature ranges used to estimate $\Theta_{\mathrm{D}}$ and has been reported in previous work \cite{sankar2017anisotropic,long2016single}. Although there is a disparity in the absolute values, both measurements indicate that $\Theta_{\mathrm{D}}$ shows only a weak dependence on $x$.

We now examine the effect of Sn doping on the specific heat of \pbsn\ in the superconducting state. The electronic contribution of the specific heat $C_{\mathrm{e}}(T)$ is determined using $C_{\mathrm{e}}(T)=C_{p}(T)-C_{\mathrm{ph}}(T)$. We plot the normalized electronic specific heat $C_{\mathrm{e}}(T)/\gamma_{\mathrm{n}}T$ versus the normalized temperature $T/T_{c}$ for different Sn concentrations in the main panel of FIG.\ref{fig:ce}. The nearly identical specific heat data observed in undoped samples with RRR = 127 and 209 strongly suggest that the presence of a minor impurity phase, as indicated in the x-ray diffraction pattern [FIG.\ref{fig:xrd}(a)], does not significantly impact the specific heat. We observe a sharp jump at $T_{c}$ for each sample. The normalized values of the specific heat jump $\Delta C/\gamma_{\mathrm{n}}T_{c}$ at $T_{c}$ are shown in FIG.\ref{fig:tc}(d) (blue squares). $\Delta C/\gamma_{\mathrm{n}}T_{c}$ is 1.71 for $x=0$ (RRR = 127) and 1.76 for $x=0$ (RRR = 209), both of which are notably larger than the predicted value of 1.43 from the BCS theory for a weak-coupling superconductor. This suggests a possible strong-coupling superconductivity realized in \pb. Our values agree with a previously reported value \cite{zhang2016superconducting}, while other studies have also reported values close to the BCS prediction \cite{sankar2017anisotropic,sun2020fully,ali2014noncentrosymmetric}. Interstingly, while slight Sn doping ($x= 0.018$) does not affect the large specific heat jump at $T_{c}$, further Sn doping suppresses $\Delta C/\gamma_{\mathrm{n}}T_{c}$ less than the predicted value for a weak-coupling superconductor.

To elucidate the superconducting gap amplitude for \pbsn, we compare our experimental data with the theoretical calculations based on the $\alpha$ model \cite{bouquet2001phenomenological}. In the $\alpha$ model, the entropy $S$ and the specific heat $C$ of the BCS weak-coupling limit can be calculated using the following equations :
\begin{equation}
\frac{S}{\gamma_{\mathrm{n}}T_{c}}=-\frac{6}{\pi^{2}}\frac{\Delta_{0}}{k_{\mathrm{B}}T_{c}}\int_{0}^{\infty}[f\ln{f}+(1-f)\ln{(1-f)}]dy,
\end{equation}

\begin{equation}
\frac{C}{\gamma_{\mathrm{n}}T_{c}}=t\frac{d(S/\gamma_{\mathrm{n}}T_{c})}{dt},
\end{equation}
where $f$ is the Fermi-Dirac distribution function defined as $f=[\exp{(\beta E)}+1]^{-1}$, $\beta=(k_{\mathrm{B}}T)^{-1}$, and $k_{\mathrm{B}}$ is the Boltzmann constant. The energy of quasiparticle excitations is given by $E=[\epsilon^{2}+\Delta^{2}(t)]^{1/2}$, where $\epsilon$ is the energy of the normal electrons relative to the Fermi energy, $t=T/T_{c}$ is the reduced temperature. The temperature dependence of the energy gap is assumed to be $\Delta (T)=\Delta_{0}\tanh{1.82[1.018(T_{c}/T-1)]^{0.51}}$, where $\Delta_{0}$ is the gap value at $T$ = 0 \cite{carrington2003magnetic,khasanov2008nodeless}. The integration variable is $y=\epsilon/\Delta_{0}$. Set $\Delta_{0}$ to be a fitting parameter, we fit the normalized electronic specific heat for $x=0$ to Eq.(4). As shown in FIGs.\ref{fig:gap}(a) and (b), the calculated curves are in excellent agreement with our data for both samples in the measured temperature range. The extracted superconducting gap amplitude, $2\Delta_{0}/k_{\mathrm{B}}T_{c}$, is 3.9 for $x=0$ (RRR = 209) and 3.8 for $x=0$ (RRR = 127), which are larger than 3.53 predicted for the BCS weak-coupling superconductors, indicating the strong-coupling superconductivity realized in \ce{PbTaSe2}. This strong-coupling superconductivity is insusceptible to slight Sn doping ($x=0.018$), as shown in FIG.\ref{fig:gap}(c).

However, the single-gap $\alpha$ model cannot reproduce the observed specific heat for moderately Sn-doped \pb. The observed specific heat jumps at $T_{c}$ for $x=0.08$ and 0.15 are 1.0 and 1.3, respectively, much smaller than the BCS weak-coupling limit of 1.43, while the specific heat of slightly Sn-doped \pb\ ($x=0.018$) is identical to undoped one, as shown in FIG.\ref{fig:ce}. We calculate the normalized specific heat using the single-gap $\alpha$ model to reproduce the observed small jumps at $T_{c}$ for $x=0.08$ and 0.15. This yields the superconducting gap amplitude $2\Delta_{0}/k_{\mathrm{B}}T_{c} = 2.9$ for $x=0.08$ and $2\Delta_{0}/k_{\mathrm{B}}T_{c} = 3.3$ for $x=0.15$, both of which are smaller than the BCS weak-coupling value of 3.53. As shown in FIGs.\ref{fig:gap}(d) and (e), the calculated curves with these small gap amplitudes (blue dashed lines) significantly deviate from the measured data. The observed smaller jumps in specific heat cannot be attributed to the presence of a nonsuperconducting fraction. To reproduce a jump size similar to that of the undoped sample, a significant nonsuperconducting fraction is needed—42\% for x = 0.08 and 26\% for x = 0.15. However, no such substantial impurity phase is detected in our x-ray diffraction measurements.

The two-gap model proposed for the multigap superconductor \ce{MgB2} explains the observed specific heat for moderately Sn-doped \pb\ within the measured temperature range \cite{bouquet2001phenomenological}. In this model, the specific heat can be expressed as $C=wC_{1}+(1-w)C_{2}$, where $C_{1}$ and $C_{2}$ represent the specific heat contribution from band 1 and 2, respectively. These contributions are calculated independently using Eqs.(3) and (4). The relative weights for each band are given by $w=\gamma_{1}/\gamma_{\mathrm{n}}$ and $1-w=\gamma_{2}/\gamma_{\mathrm{n}}$, where $\gamma_{i}$ is a partial Sommerfeld constant for band $i$ and $\gamma_{\mathrm{n}}=\gamma_{1}+\gamma_{2}$. As shown in FIG.\ref{fig:gap} (d) and (e), the theoretical calculations based on the two-gap model are in excellent agreement with our data across the measured temperature range. We obtain the parameters $2\Delta_{1}/k_{\mathrm{B}}T_{c}=3.9$, $2\Delta_{2}/k_{\mathrm{B}}T_{c}=0.91$, and $\gamma_{1}:\gamma_{2}=55:45$ for $x=0.080$ and $2\Delta_{1}/k_{\mathrm{B}}T_{c}=3.9$, $2\Delta_{2}/k_{\mathrm{B}}T_{c}=0.89$, and $\gamma_{1}:\gamma_{2}=72:28$ for $x=0.15$. The larger gap amplitudes of $x=0.080$ and 0.15 are very close to those of $x=0$ and 0.018, and appear to be independent of Sn doping. The smaller gap amplitude of $x=0.080$ is comparable to that of $x=0.15$ and seems nearly independent of Sn doping. On the other hand, the ratio of the density of states for bands 1 and 2 varies with Sn doping, which affects the overall shape of the specific heat. Our observations indicate that moderate Sn doping significantly impacts the multigap nature in the superconducting state of \pbsn.

We note that the data for undoped and slightly doped samples can be fitted using the two-gap model, in which the gap sizes are fixed as follows: $2\Delta_{1}/k_{\mathrm{B}}T_{c}=3.9$ and $2\Delta_{2}/k_{\mathrm{B}}T_{c}=0.90$ for the undoped sample with RRR = 209, and $2\Delta_{1}/k_{\mathrm{B}}T_{c}=3.8$ and $2\Delta_{2}/k_{\mathrm{B}}T_{c}=0.90$ for the undoped sample with RRR = 127 and the slightly Sn-doped sample ($x=0.018$). The relative contribution of band 1 to the total density of states is used as a fitting parameter. These fittings yield a ratio of $\gamma_{1}:\gamma_{2}=96:4$, consistent with the values used to describe the superfluid density of undoped \ce{PbTaSe2} obtained from $\mu$SR measurements \cite{wilson2017mu}. Although both the single-gap and two-gap fittings reproduce the data equally well, the contribution from the passive band in the two-gap model is negligibly small. These results suggest that multiband superconductivity plays a minor role in undoped and slightly doped \ce{PbTaSe2}.

We observe an enhancement of $T_{c}$ with Sn doping in the noncentrosymmetric superconductor \pbsn, where odd-parity superconducting states are theoretically allowed to exist. This enhancement is seemingly counterintuitive because the impurity scattering due to disorder increases significantly---by a factor of 50, as estimated from $\rho_{0}$ and by a factor of 20 from the RRR---and in general, disorder negatively affects the superconductivity. This suggests that superconductivity is strikingly robust against disorders, contrasting with odd-parity superconductivity. The robustness against disorders is further supported by our observations of no discernible effect on the specific heat and $T_{c}$ in undoped \pb\ with two different RRR values. These results imply that the contribution from odd-parity pairing in the superconducting gap function is negligible in \pb\, consistent with previous work reporting the fully-gapped conventional superconducting state \cite{zhang2016superconducting,wang16,pang16,wilson2017mu}. 

Qualitatively, substituting Pb with Sn can enhance $T_{c}$ due to the lighter atomic mass $M$ of Sn compared to Pb. This substitution slightly increases the Debye temperature ($\propto M^{-1/2}$) as well as the electron-phonon coupling strength $\lambda$ ($\propto M^{-1}$), leading to an enhancement of $T_{c}$, as described by the McMillan formula,
\begin{equation}
  T_{c}=\frac{\Theta_{\mathrm{D}}}{1.45}\exp\left [-\frac{1.04(1+\lambda)}{\lambda-\mu^{\ast}(1+0.62\lambda)}\right],
\end{equation}
where $\mu^{\ast}=0.10-0.15$ is the Coulomb repulsive screened parameter \cite{mcmil68}. However, the modest increase in $\Theta_{\mathrm{D}}$ or $\lambda$ resulting from $\sim$ 20 \% Sn doping is likely too small to account for the enhancement of $T_{c}$ by more than 40\%. The enhancement of effective electron-phonon coupling strength estimated from experimental values using Eq.(5) is discernibly larger than expected solely from the atomic mass contribution. By utilizing $\mu^{\ast}=0.13$, the measured bulk $T_{c}$, and $\Theta_{\mathrm{D}}^{C}$ with Eq.(5), we obtain $\lambda = 0.73$ for $x=0$ and $\lambda = 0.80$ for $x=0.15$, an increase by $\sim$ 10\%. On the other hand, the expected enhancement of $\lambda$ due to the atomic mass alone can be estimated from the ratio of effective atomic masses \cite{bouvi91}:
\begin{equation}
\begin{split}
  \frac{\lambda(\ce{Pb_{1-x}Sn_{x}TaSe2})}{\lambda(\ce{PbTaSe2})}\sim\left[ \frac{\Theta_{D}(\ce{Pb_{1-x}Sn_{x}TaSe2})}{\Theta_{D}(\ce{PbTaSe2})} \right]^{2}\\
  =\left[\frac{M_{\ce{Pb}}^{3/2}+M_{\ce{Ta}}^{3/2}+2M_{\ce{Se}}^{3/2}}{(1-x)M_{\ce{Pb}}^{3/2}+xM_{\ce{Sn}}^{3/2}+M_{\ce{Ta}}^{3/2}+2M_{\ce{Se}}^{3/2}}\right]^{2/3},
 \end{split}
 \end{equation}
yielding $\lambda(\ce{Pb_{0.85}Sn_{0.15}TaSe2})/\lambda(\ce{PbTaSe2})= 1.03$. This discrepancy between the observed and expected enhancement of $\lambda$ indicates additional contributions to $\lambda$ other than the atomic mass effect. 

We attribute this enhancement of the electron-phonon coupling strength to the multiband effect induced by Sn doping, which controls the spin-orbit gap in the band structure of \pbsn. The specific heat data for $x=0.080$ and $0.15$ necessitate the two-gap model to account for the observations, indicating that the multiband effect plays an essential role in the superconducting states of \pbsn.

\begin{figure*}[htb]
\includegraphics[width=0.9\textwidth]{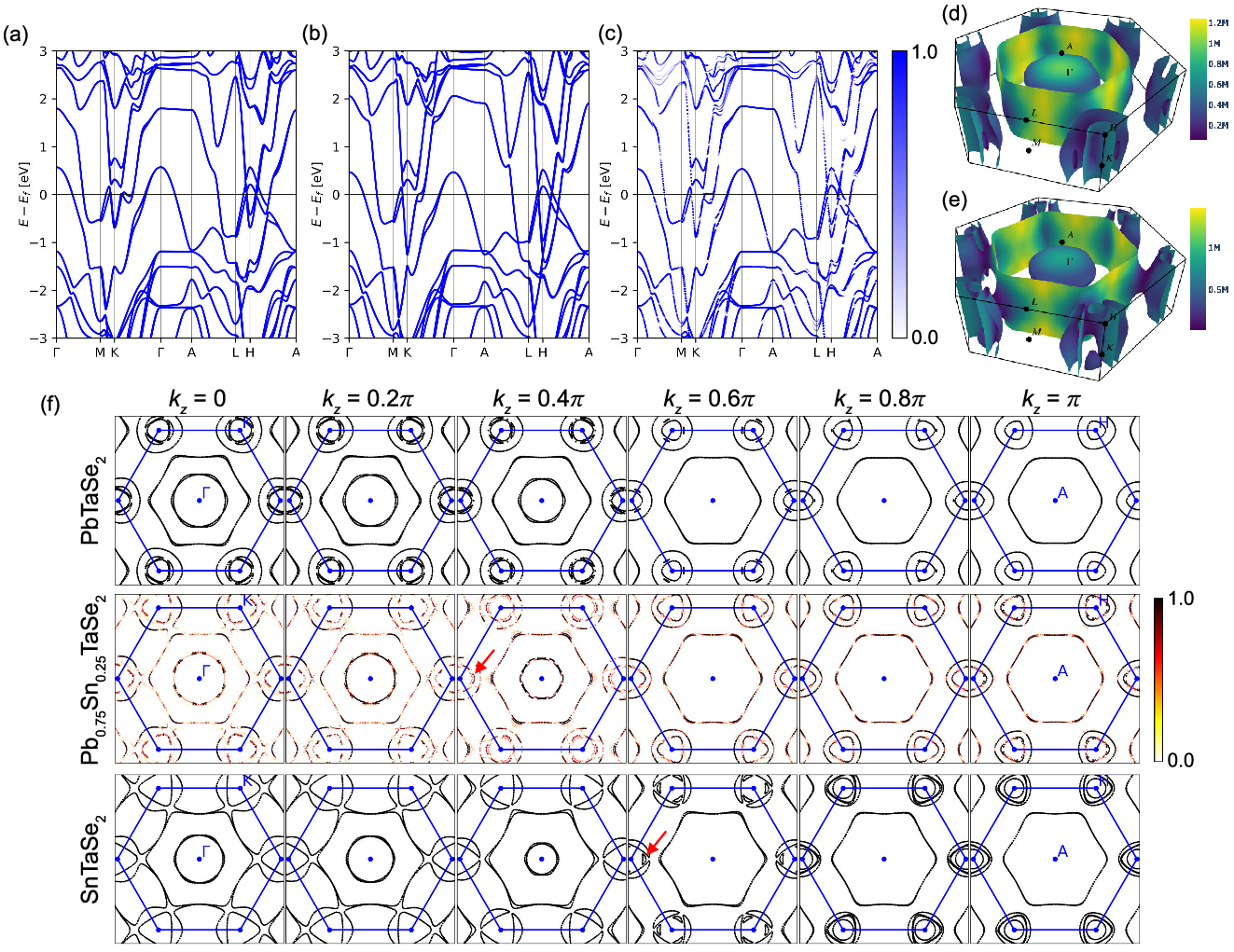}
\caption{\label{fig:fs}Band structures of (a) \ce{PbTaSe2} and (b) \ce{SnTaSe2}. (c) Unfolded band structure of \ce{Pb_{0.75}Sn_{0.25}TaSe2}. The color scale represents the spectral weights. Fermi surfaces for (d) \ce{PbTaSe2} and (e) \ce{SnTaSe2}. The color scale represents the Fermi velocity in m/s. (f) Two dimentional slices of Fermi surfaces of \ce{PbTaSe2} (top), \ce{Pb_{0.75}Sn_{0.25}TaSe2} (middle), and \ce{SnTaSe2} (bottom). The color scale for \ce{Pb_{0.75}Sn_{0.25}TaSe2} represents the spectral weights. Red arrows indicate the crescent-shaped cross section of the trigonal-bipyramid-like Fermi pockets near the $H$ points as observed in \ce{SnTaSe2}.}
\end{figure*}

The multiband effect arises from changes in the shapes of the Fermi surfaces induced by Sn doping. As shown in FIGs. \ref{fig:fs}(a) and (b), the calculated band structures for \ce{PbTaSe2} and \ce{SnTaSe2} appear quite similar, consistent with the previous \emph{ab initio} calculations \cite{chen16a}. However, subtle differences in the band structures near the $K$ and $H$ points lead to significant alterations in the Fermi surface shapes around these regions, which have not been pointed out in the previous work. In \ce{PbTaSe2}, Fermi surfaces near the $K$-$H$ lines are quasi-two dimensional, whereas in \ce{SnTaSe2}, they become distinctly three dimensional. Trigonal-bipyramid-like Fermi pockets emerge near the $H$ points, and compressed tubular Fermi surfaces appear near the $K$-$H$ lines, connecting to a large hexagonal cylindrical Fermi surface centered at the $\Gamma$ point.

This evolution in dimensionality near the $K$ and $H$ points must begin at a doping level $x< 0.25$, as we observe clear evidence of three-dimensional trigonal-bipyramid-like Fermi pockets already at $x=0.25$. As illustrated in FIG. \ref{fig:fs}(f), the two-dimentional slices of the Fermi surfaces of \ce{Pb_{0.75}Sn_{0.25}TaSe2} more closely resemble those of \ce{SnTaSe2} than those of \ce{PbTaSe2}. Notably, we observe the emergence of trigonal-bipyramid-like Fermi pockets near the $H$ points at $x=0.25$, evidenced by the crescent-shaped cross section (indicated by a red arrow) in the slice at $k_{z} = 0.4\pi$---similar to the feature observed at $k_{z}=0.6\pi$ in \ce{SnTaSe2}. These findings suggest that the quasi-two-dimensional Fermi surfaces in \ce{PbTaSe2} around the $K$-$H$ lines begin to split at $k_{z}=0$ for $x< 0.25$, evolving into fully three-dimensional pockets as the Sn concentration increases.

The emergence of additional three-dimensional Fermi surfaces induced by Sn doping activates the multiband superconductivity in \pbsn. In this scenario, activating the interband coupling between bands enhances the effective electron-phonon coupling constant, inevitably increasing $T_{c}$ \cite{henhe24}, while the larger gap size $2\Delta_{1}/k_{\mathrm{B}}T_{c}$ remains unaffected by Sn doping. In addition, the electron-phonon coupling of \pbsn\ can be enhanced by the quantum geometric effect due to nontrivial electron band geometry and/or topology \cite{yu24}. Indeed, the nontrivial aspect of the band topology of \ce{PbTaSe2} can remain as Sn is substituted for Pb \cite{chen16a}. Further experimental and theoretical efforts are needed to verify this quantum geometric contribution in \pbsn.

\section{Summary}
In summary, we have measured the resistivity and specific heat of \pbsn\ single crystals. We observe the enhancement of $T_{c}$ with the increasing Sn concentration $x$ while also finding a significant increase in impurity scattering with Sn doping, as indicated by the residual resistivity and the residual resistivity ratio. For undoped ($x=0$) and slightly doped ($x=0.018$) samples, the specific heat jump at $T_{c}$ is greater than the BCS value of 1.43, yet the jump diminishes with Sn doping. The specific heat data of moderately Sn-doped samples are reproduced by theoretical calculations based on the two-gap model, highlighting the pivotal role of emergent three-dimensional Fermi pockets caused by the Sn substitution in the superconducting state. The enhancement of $T_{c}$ in \pbsn\ can be attributed to the multiband effect due to these emergent Fermi pockets, driven by a reduction of the spin-orbit coupling via Sn doping.

\begin{acknowledgments}
The authors thank V. Turkowski for his valuable discussions. This work was supported by an NSF Career DMR-1944975. DL acknowledges funding from the US Department of Energy under grant DE-FG02-07ER46354. Computational resource was provided by the University of Central Florida's Advanced Research Computing Center.
\end{acknowledgments}

\section*{Data Availability}
The data that support the findings of this article are openly available \cite{da}.


\begin{thebibliography}{39}%
\makeatletter
\providecommand \@ifxundefined [1]{%
 \@ifx{#1\undefined}
}%
\providecommand \@ifnum [1]{%
 \ifnum #1\expandafter \@firstoftwo
 \else \expandafter \@secondoftwo
 \fi
}%
\providecommand \@ifx [1]{%
 \ifx #1\expandafter \@firstoftwo
 \else \expandafter \@secondoftwo
 \fi
}%
\providecommand \natexlab [1]{#1}%
\providecommand \enquote  [1]{``#1''}%
\providecommand \bibnamefont  [1]{#1}%
\providecommand \bibfnamefont [1]{#1}%
\providecommand \citenamefont [1]{#1}%
\providecommand \href@noop [0]{\@secondoftwo}%
\providecommand \href [0]{\begingroup \@sanitize@url \@href}%
\providecommand \@href[1]{\@@startlink{#1}\@@href}%
\providecommand \@@href[1]{\endgroup#1\@@endlink}%
\providecommand \@sanitize@url [0]{\catcode `\\12\catcode `\$12\catcode
  `\&12\catcode `\#12\catcode `\^12\catcode `\_12\catcode `\%12\relax}%
\providecommand \@@startlink[1]{}%
\providecommand \@@endlink[0]{}%
\providecommand \url  [0]{\begingroup\@sanitize@url \@url }%
\providecommand \@url [1]{\endgroup\@href {#1}{\urlprefix }}%
\providecommand \urlprefix  [0]{URL }%
\providecommand \Eprint [0]{\href }%
\providecommand \doibase [0]{http://dx.doi.org/}%
\providecommand \selectlanguage [0]{\@gobble}%
\providecommand \bibinfo  [0]{\@secondoftwo}%
\providecommand \bibfield  [0]{\@secondoftwo}%
\providecommand \translation [1]{[#1]}%
\providecommand \BibitemOpen [0]{}%
\providecommand \bibitemStop [0]{}%
\providecommand \bibitemNoStop [0]{.\EOS\space}%
\providecommand \EOS [0]{\spacefactor3000\relax}%
\providecommand \BibitemShut  [1]{\csname bibitem#1\endcsname}%
\let\auto@bib@innerbib\@empty
\bibitem [{\citenamefont {Bauer}\ \emph {et~al.}(2004)\citenamefont {Bauer},
  \citenamefont {Hilscher}, \citenamefont {Michor}, \citenamefont {Paul},
  \citenamefont {Scheidt}, \citenamefont {Gribanov}, \citenamefont {Seropegin},
  \citenamefont {No\"el}, \citenamefont {Sigrist},\ and\ \citenamefont
  {Rogl}}]{bauer04}%
  \BibitemOpen
  \bibfield  {author} {\bibinfo {author} {\bibfnamefont {E.}~\bibnamefont
  {Bauer}}, \bibinfo {author} {\bibfnamefont {G.}~\bibnamefont {Hilscher}},
  \bibinfo {author} {\bibfnamefont {H.}~\bibnamefont {Michor}}, \bibinfo
  {author} {\bibfnamefont {C.}~\bibnamefont {Paul}}, \bibinfo {author}
  {\bibfnamefont {E.~W.}\ \bibnamefont {Scheidt}}, \bibinfo {author}
  {\bibfnamefont {A.}~\bibnamefont {Gribanov}}, \bibinfo {author}
  {\bibfnamefont {Y.}~\bibnamefont {Seropegin}}, \bibinfo {author}
  {\bibfnamefont {H.}~\bibnamefont {No\"el}}, \bibinfo {author} {\bibfnamefont
  {M.}~\bibnamefont {Sigrist}}, \ and\ \bibinfo {author} {\bibfnamefont
  {P.}~\bibnamefont {Rogl}},\ }\bibfield  {title} {\bibinfo {title} {Heavy
  Fermion Superconductivity and Magnetic Order in Noncentrosymmetric
  ${\mathrm{C}\mathrm{e}\mathrm{P}\mathrm{t}}_{3}\mathrm{S}\mathrm{i}$},\
  }\href {\doibase 10.1103/PhysRevLett.92.027003} {\bibfield  {journal}
  {\bibinfo  {journal} {Phys. Rev. Lett.}\ }\textbf {\bibinfo {volume} {92}},\
  \bibinfo {pages} {027003} (\bibinfo {year} {2004})}\BibitemShut {NoStop}%
\bibitem [{\citenamefont {Kimura}\ \emph {et~al.}(2007)\citenamefont {Kimura},
  \citenamefont {Ito}, \citenamefont {Aoki}, \citenamefont {Uji},\ and\
  \citenamefont {Terashima}}]{kimur07}%
  \BibitemOpen
  \bibfield  {author} {\bibinfo {author} {\bibfnamefont {N.}~\bibnamefont
  {Kimura}}, \bibinfo {author} {\bibfnamefont {K.}~\bibnamefont {Ito}},
  \bibinfo {author} {\bibfnamefont {H.}~\bibnamefont {Aoki}}, \bibinfo {author}
  {\bibfnamefont {S.}~\bibnamefont {Uji}}, \ and\ \bibinfo {author}
  {\bibfnamefont {T.}~\bibnamefont {Terashima}},\ }\bibfield  {title} {\bibinfo
  {title} {Extremely High Upper Critical Magnetic Field of the
  Noncentrosymmetric Heavy Fermion Superconductor ${\mathrm{CeRhSi}}_{3}$},\
  }\href {\doibase 10.1103/PhysRevLett.98.197001} {\bibfield  {journal}
  {\bibinfo  {journal} {Phys. Rev. Lett.}\ }\textbf {\bibinfo {volume} {98}},\
  \bibinfo {pages} {197001} (\bibinfo {year} {2007})}\BibitemShut {NoStop}%
\bibitem [{\citenamefont {Yuan}\ \emph {et~al.}(2006)\citenamefont {Yuan},
  \citenamefont {Agterberg}, \citenamefont {Hayashi}, \citenamefont {Badica},
  \citenamefont {Vandervelde}, \citenamefont {Togano}, \citenamefont
  {Sigrist},\ and\ \citenamefont {Salamon}}]{yuan06}%
  \BibitemOpen
  \bibfield  {author} {\bibinfo {author} {\bibfnamefont {H.~Q.}\ \bibnamefont
  {Yuan}}, \bibinfo {author} {\bibfnamefont {D.~F.}\ \bibnamefont {Agterberg}},
  \bibinfo {author} {\bibfnamefont {N.}~\bibnamefont {Hayashi}}, \bibinfo
  {author} {\bibfnamefont {P.}~\bibnamefont {Badica}}, \bibinfo {author}
  {\bibfnamefont {D.}~\bibnamefont {Vandervelde}}, \bibinfo {author}
  {\bibfnamefont {K.}~\bibnamefont {Togano}}, \bibinfo {author} {\bibfnamefont
  {M.}~\bibnamefont {Sigrist}}, \ and\ \bibinfo {author} {\bibfnamefont
  {M.~B.}\ \bibnamefont {Salamon}},\ }\bibfield  {title} {\bibinfo {title}
  {$S$-Wave Spin-Triplet Order in Superconductors without Inversion Symmetry:
  ${\mathrm{Li}}_{2}{\mathrm{Pd}}_{3}\mathrm{B}$ and
  ${\mathrm{Li}}_{2}{\mathrm{Pt}}_{3}\mathrm{B}$},\ }\href {\doibase
  10.1103/PhysRevLett.97.017006} {\bibfield  {journal} {\bibinfo  {journal}
  {Phys. Rev. Lett.}\ }\textbf {\bibinfo {volume} {97}},\ \bibinfo {pages}
  {017006} (\bibinfo {year} {2006})}\BibitemShut {NoStop}%
\bibitem [{\citenamefont {Nishiyama}\ \emph {et~al.}(2007)\citenamefont
  {Nishiyama}, \citenamefont {Inada},\ and\ \citenamefont {Zheng}}]{nishi07}%
  \BibitemOpen
  \bibfield  {author} {\bibinfo {author} {\bibfnamefont {M.}~\bibnamefont
  {Nishiyama}}, \bibinfo {author} {\bibfnamefont {Y.}~\bibnamefont {Inada}}, \
  and\ \bibinfo {author} {\bibfnamefont {G.-q.}\ \bibnamefont {Zheng}},\
  }\bibfield  {title} {\bibinfo {title} {Spin Triplet Superconducting State due
  to Broken Inversion Symmetry in
  ${\mathrm{Li}}_{2}{\mathrm{Pt}}_{3}\mathrm{B}$},\ }\href {\doibase
  10.1103/PhysRevLett.98.047002} {\bibfield  {journal} {\bibinfo  {journal}
  {Phys. Rev. Lett.}\ }\textbf {\bibinfo {volume} {98}},\ \bibinfo {pages}
  {047002} (\bibinfo {year} {2007})}\BibitemShut {NoStop}%
\bibitem [{\citenamefont {Chen}\ \emph {et~al.}(2011)\citenamefont {Chen},
  \citenamefont {Salamon}, \citenamefont {Akutagawa}, \citenamefont {Akimitsu},
  \citenamefont {Singleton}, \citenamefont {Zhang}, \citenamefont {Jiao},\ and\
  \citenamefont {Yuan}}]{chen11}%
  \BibitemOpen
  \bibfield  {author} {\bibinfo {author} {\bibfnamefont {J.}~\bibnamefont
  {Chen}}, \bibinfo {author} {\bibfnamefont {M.~B.}\ \bibnamefont {Salamon}},
  \bibinfo {author} {\bibfnamefont {S.}~\bibnamefont {Akutagawa}}, \bibinfo
  {author} {\bibfnamefont {J.}~\bibnamefont {Akimitsu}}, \bibinfo {author}
  {\bibfnamefont {J.}~\bibnamefont {Singleton}}, \bibinfo {author}
  {\bibfnamefont {J.~L.}\ \bibnamefont {Zhang}}, \bibinfo {author}
  {\bibfnamefont {L.}~\bibnamefont {Jiao}}, \ and\ \bibinfo {author}
  {\bibfnamefont {H.~Q.}\ \bibnamefont {Yuan}},\ }\bibfield  {title} {\bibinfo
  {title} {Evidence of nodal gap structure in the noncentrosymmetric
  superconductor ${\mathrm{Y}}_{2}$C${}_{3}$},\ }\href {\doibase
  10.1103/PhysRevB.83.144529} {\bibfield  {journal} {\bibinfo  {journal} {Phys.
  Rev. B}\ }\textbf {\bibinfo {volume} {83}},\ \bibinfo {pages} {144529}
  (\bibinfo {year} {2011})}\BibitemShut {NoStop}%
\bibitem [{\citenamefont {Harada}\ \emph {et~al.}(2012)\citenamefont {Harada},
  \citenamefont {Zhou}, \citenamefont {Yao}, \citenamefont {Inada},\ and\
  \citenamefont {Zheng}}]{harad12}%
  \BibitemOpen
  \bibfield  {author} {\bibinfo {author} {\bibfnamefont {S.}~\bibnamefont
  {Harada}}, \bibinfo {author} {\bibfnamefont {J.~J.}\ \bibnamefont {Zhou}},
  \bibinfo {author} {\bibfnamefont {Y.~G.}\ \bibnamefont {Yao}}, \bibinfo
  {author} {\bibfnamefont {Y.}~\bibnamefont {Inada}}, \ and\ \bibinfo {author}
  {\bibfnamefont {G.-q.}\ \bibnamefont {Zheng}},\ }\bibfield  {title} {\bibinfo
  {title} {Abrupt enhancement of noncentrosymmetry and appearance of a
  spin-triplet superconducting state in Li$_{2}$(Pd$_{1-x}$Pt$_{x}$)$_{3}$B
  beyond $x=0.8$},\ }\href {\doibase 10.1103/PhysRevB.86.220502} {\bibfield
  {journal} {\bibinfo  {journal} {Phys. Rev. B}\ }\textbf {\bibinfo {volume}
  {86}},\ \bibinfo {pages} {220502} (\bibinfo {year} {2012})}\BibitemShut
  {NoStop}%
\bibitem [{\citenamefont {Sato}\ and\ \citenamefont {Ando}(2017)}]{sato17}%
  \BibitemOpen
  \bibfield  {author} {\bibinfo {author} {\bibfnamefont {M.}~\bibnamefont
  {Sato}}\ and\ \bibinfo {author} {\bibfnamefont {Y.}~\bibnamefont {Ando}},\
  }\bibfield  {title} {\bibinfo {title} {Topological superconductors: a
  review},\ }\href {http://stacks.iop.org/0034-4885/80/i=7/a=076501} {\bibfield
   {journal} {\bibinfo  {journal} {Rep. Prog. Phys.}\ }\textbf {\bibinfo
  {volume} {80}},\ \bibinfo {pages} {076501} (\bibinfo {year}
  {2017})}\BibitemShut {NoStop}%
\bibitem [{\citenamefont {Zhang}\ \emph {et~al.}(2016)\citenamefont {Zhang},
  \citenamefont {Yuan}, \citenamefont {Bian}, \citenamefont {Xu}, \citenamefont
  {Zhang}, \citenamefont {Hasan},\ and\ \citenamefont
  {Jia}}]{zhang2016superconducting}%
  \BibitemOpen
  \bibfield  {author} {\bibinfo {author} {\bibfnamefont {C.-L.}\ \bibnamefont
  {Zhang}}, \bibinfo {author} {\bibfnamefont {Z.}~\bibnamefont {Yuan}},
  \bibinfo {author} {\bibfnamefont {G.}~\bibnamefont {Bian}}, \bibinfo {author}
  {\bibfnamefont {S.-Y.}\ \bibnamefont {Xu}}, \bibinfo {author} {\bibfnamefont
  {X.}~\bibnamefont {Zhang}}, \bibinfo {author} {\bibfnamefont {M.~Z.}\
  \bibnamefont {Hasan}}, \ and\ \bibinfo {author} {\bibfnamefont
  {S.}~\bibnamefont {Jia}},\ }\bibfield  {title} {\bibinfo {title}
  {Superconducting properties in single crystals of the topological nodal
  semimetal ${\mathrm{PbTaSe}}_{2}$},\ }\href@noop {} {\bibfield  {journal}
  {\bibinfo  {journal} {Phys. Rev. B}\ }\textbf {\bibinfo {volume} {93}},\
  \bibinfo {pages} {054520} (\bibinfo {year} {2016})}\BibitemShut {NoStop}%
\bibitem [{\citenamefont {Wang}\ \emph {et~al.}(2015)\citenamefont {Wang},
  \citenamefont {Xu}, \citenamefont {Zhou}, \citenamefont {Li}, \citenamefont
  {Cao}, \citenamefont {Yang}, \citenamefont {Li}, \citenamefont {Cao},
  \citenamefont {Dai}, \citenamefont {Zhang}, \citenamefont {Shi},
  \citenamefont {Chen},\ and\ \citenamefont {Yang}}]{wang15b}%
  \BibitemOpen
  \bibfield  {author} {\bibinfo {author} {\bibfnamefont {J.}~\bibnamefont
  {Wang}}, \bibinfo {author} {\bibfnamefont {X.}~\bibnamefont {Xu}}, \bibinfo
  {author} {\bibfnamefont {N.}~\bibnamefont {Zhou}}, \bibinfo {author}
  {\bibfnamefont {L.}~\bibnamefont {Li}}, \bibinfo {author} {\bibfnamefont
  {X.}~\bibnamefont {Cao}}, \bibinfo {author} {\bibfnamefont {J.}~\bibnamefont
  {Yang}}, \bibinfo {author} {\bibfnamefont {Y.}~\bibnamefont {Li}}, \bibinfo
  {author} {\bibfnamefont {C.}~\bibnamefont {Cao}}, \bibinfo {author}
  {\bibfnamefont {J.}~\bibnamefont {Dai}}, \bibinfo {author} {\bibfnamefont
  {J.}~\bibnamefont {Zhang}}, \bibinfo {author} {\bibfnamefont
  {Z.}~\bibnamefont {Shi}}, \bibinfo {author} {\bibfnamefont {B.}~\bibnamefont
  {Chen}}, \ and\ \bibinfo {author} {\bibfnamefont {Z.}~\bibnamefont {Yang}},\
  }\bibfield  {title} {\bibinfo {title} {Upward Curvature of the Upper Critical
  Field and the V-Shaped Pressure Dependence of Tcin the Noncentrosymmetric
  Superconductor ${\mathrm{PbTaSe}}_{2}$},\ }\href {\doibase
  10.1007/s10948-015-3177-4} {\bibfield  {journal} {\bibinfo  {journal} {J.
  Supercond. Nov. Magn.}\ }\textbf {\bibinfo {volume} {28}},\ \bibinfo {pages}
  {3173} (\bibinfo {year} {2015})}\BibitemShut {NoStop}%
\bibitem [{\citenamefont {Ali}\ \emph {et~al.}(2014)\citenamefont {Ali},
  \citenamefont {Gibson}, \citenamefont {Klimczuk},\ and\ \citenamefont
  {Cava}}]{ali2014noncentrosymmetric}%
  \BibitemOpen
  \bibfield  {author} {\bibinfo {author} {\bibfnamefont {M.~N.}\ \bibnamefont
  {Ali}}, \bibinfo {author} {\bibfnamefont {Q.~D.}\ \bibnamefont {Gibson}},
  \bibinfo {author} {\bibfnamefont {T.}~\bibnamefont {Klimczuk}}, \ and\
  \bibinfo {author} {\bibfnamefont {R.~J.}\ \bibnamefont {Cava}},\ }\bibfield
  {title} {\bibinfo {title} {Noncentrosymmetric superconductor with a bulk
  three-dimensional Dirac cone gapped by strong spin-orbit coupling},\
  }\href@noop {} {\bibfield  {journal} {\bibinfo  {journal} {Phys. Rev. B}\
  }\textbf {\bibinfo {volume} {89}},\ \bibinfo {pages} {020505} (\bibinfo
  {year} {2014})}\BibitemShut {NoStop}%
\bibitem [{\citenamefont {Wang}\ \emph {et~al.}(2016)\citenamefont {Wang},
  \citenamefont {Xu}, \citenamefont {He}, \citenamefont {Zhang}, \citenamefont
  {Hong}, \citenamefont {Cai}, \citenamefont {Wang}, \citenamefont {Dong},\
  and\ \citenamefont {Li}}]{wang16}%
  \BibitemOpen
  \bibfield  {author} {\bibinfo {author} {\bibfnamefont {M.~X.}\ \bibnamefont
  {Wang}}, \bibinfo {author} {\bibfnamefont {Y.}~\bibnamefont {Xu}}, \bibinfo
  {author} {\bibfnamefont {L.~P.}\ \bibnamefont {He}}, \bibinfo {author}
  {\bibfnamefont {J.}~\bibnamefont {Zhang}}, \bibinfo {author} {\bibfnamefont
  {X.~C.}\ \bibnamefont {Hong}}, \bibinfo {author} {\bibfnamefont {P.~L.}\
  \bibnamefont {Cai}}, \bibinfo {author} {\bibfnamefont {Z.~B.}\ \bibnamefont
  {Wang}}, \bibinfo {author} {\bibfnamefont {J.~K.}\ \bibnamefont {Dong}}, \
  and\ \bibinfo {author} {\bibfnamefont {S.~Y.}\ \bibnamefont {Li}},\
  }\bibfield  {title} {\bibinfo {title} {Nodeless superconducting gaps in
  noncentrosymmetric superconductor ${\mathrm{PbTaSe}}_{2}$ with topological
  bulk nodal lines},\ }\href {\doibase 10.1103/PhysRevB.93.020503} {\bibfield
  {journal} {\bibinfo  {journal} {Phys. Rev. B}\ }\textbf {\bibinfo {volume}
  {93}},\ \bibinfo {pages} {020503} (\bibinfo {year} {2016})}\BibitemShut
  {NoStop}%
\bibitem [{\citenamefont {Pang}\ \emph {et~al.}(2016)\citenamefont {Pang},
  \citenamefont {Smidman}, \citenamefont {Zhao}, \citenamefont {Wang},
  \citenamefont {Weng}, \citenamefont {Che}, \citenamefont {Chen},
  \citenamefont {Lu}, \citenamefont {Chen},\ and\ \citenamefont
  {Yuan}}]{pang16}%
  \BibitemOpen
  \bibfield  {author} {\bibinfo {author} {\bibfnamefont {G.~M.}\ \bibnamefont
  {Pang}}, \bibinfo {author} {\bibfnamefont {M.}~\bibnamefont {Smidman}},
  \bibinfo {author} {\bibfnamefont {L.~X.}\ \bibnamefont {Zhao}}, \bibinfo
  {author} {\bibfnamefont {Y.~F.}\ \bibnamefont {Wang}}, \bibinfo {author}
  {\bibfnamefont {Z.~F.}\ \bibnamefont {Weng}}, \bibinfo {author}
  {\bibfnamefont {L.~Q.}\ \bibnamefont {Che}}, \bibinfo {author} {\bibfnamefont
  {Y.}~\bibnamefont {Chen}}, \bibinfo {author} {\bibfnamefont {X.}~\bibnamefont
  {Lu}}, \bibinfo {author} {\bibfnamefont {G.~F.}\ \bibnamefont {Chen}}, \ and\
  \bibinfo {author} {\bibfnamefont {H.~Q.}\ \bibnamefont {Yuan}},\ }\bibfield
  {title} {\bibinfo {title} {Nodeless superconductivity in noncentrosymmetric
  ${\mathrm{PbTaSe}}_{2}$ single crystals},\ }\href {\doibase
  10.1103/PhysRevB.93.060506} {\bibfield  {journal} {\bibinfo  {journal} {Phys.
  Rev. B}\ }\textbf {\bibinfo {volume} {93}},\ \bibinfo {pages} {060506}
  (\bibinfo {year} {2016})}\BibitemShut {NoStop}%
\bibitem [{\citenamefont {Wilson}\ \emph {et~al.}(2017)\citenamefont {Wilson},
  \citenamefont {Hallas}, \citenamefont {Cai}, \citenamefont {Guo},
  \citenamefont {Gong}, \citenamefont {Sankar}, \citenamefont {Chou},
  \citenamefont {Uemura},\ and\ \citenamefont {Luke}}]{wilson2017mu}%
  \BibitemOpen
  \bibfield  {author} {\bibinfo {author} {\bibfnamefont {M.~N.}\ \bibnamefont
  {Wilson}}, \bibinfo {author} {\bibfnamefont {A.~M.}\ \bibnamefont {Hallas}},
  \bibinfo {author} {\bibfnamefont {Y.}~\bibnamefont {Cai}}, \bibinfo {author}
  {\bibfnamefont {S.}~\bibnamefont {Guo}}, \bibinfo {author} {\bibfnamefont
  {Z.}~\bibnamefont {Gong}}, \bibinfo {author} {\bibfnamefont {R.}~\bibnamefont
  {Sankar}}, \bibinfo {author} {\bibfnamefont {F.~C.}\ \bibnamefont {Chou}},
  \bibinfo {author} {\bibfnamefont {Y.~J.}\ \bibnamefont {Uemura}}, \ and\
  \bibinfo {author} {\bibfnamefont {G.~M.}\ \bibnamefont {Luke}},\ }\bibfield
  {title} {\bibinfo {title} {$\ensuremath{\mu}\mathrm{SR}$ study of the
  noncentrosymmetric superconductor ${\mathrm{PbTaSe}}_{2}$},\ }\href {\doibase
  10.1103/PhysRevB.95.224506} {\bibfield  {journal} {\bibinfo  {journal} {Phys.
  Rev. B}\ }\textbf {\bibinfo {volume} {95}},\ \bibinfo {pages} {224506}
  (\bibinfo {year} {2017})}\BibitemShut {NoStop}%
\bibitem [{\citenamefont {Chang}\ \emph {et~al.}(2016)\citenamefont {Chang},
  \citenamefont {Chen}, \citenamefont {Bian}, \citenamefont {Huang},
  \citenamefont {Zheng}, \citenamefont {Neupert}, \citenamefont {Sankar},
  \citenamefont {Xu}, \citenamefont {Belopolski}, \citenamefont {Chang},
  \citenamefont {Wang}, \citenamefont {Chou}, \citenamefont {Bansil},
  \citenamefont {Jeng}, \citenamefont {Lin},\ and\ \citenamefont
  {Hasan}}]{chang16}%
  \BibitemOpen
  \bibfield  {author} {\bibinfo {author} {\bibfnamefont {T.-R.}\ \bibnamefont
  {Chang}}, \bibinfo {author} {\bibfnamefont {P.-J.}\ \bibnamefont {Chen}},
  \bibinfo {author} {\bibfnamefont {G.}~\bibnamefont {Bian}}, \bibinfo {author}
  {\bibfnamefont {S.-M.}\ \bibnamefont {Huang}}, \bibinfo {author}
  {\bibfnamefont {H.}~\bibnamefont {Zheng}}, \bibinfo {author} {\bibfnamefont
  {T.}~\bibnamefont {Neupert}}, \bibinfo {author} {\bibfnamefont
  {R.}~\bibnamefont {Sankar}}, \bibinfo {author} {\bibfnamefont {S.-Y.}\
  \bibnamefont {Xu}}, \bibinfo {author} {\bibfnamefont {I.}~\bibnamefont
  {Belopolski}}, \bibinfo {author} {\bibfnamefont {G.}~\bibnamefont {Chang}},
  \bibinfo {author} {\bibfnamefont {B.}~\bibnamefont {Wang}}, \bibinfo {author}
  {\bibfnamefont {F.}~\bibnamefont {Chou}}, \bibinfo {author} {\bibfnamefont
  {A.}~\bibnamefont {Bansil}}, \bibinfo {author} {\bibfnamefont {H.-T.}\
  \bibnamefont {Jeng}}, \bibinfo {author} {\bibfnamefont {H.}~\bibnamefont
  {Lin}}, \ and\ \bibinfo {author} {\bibfnamefont {M.~Z.}\ \bibnamefont
  {Hasan}},\ }\bibfield  {title} {\bibinfo {title} {Topological Dirac surface
  states and superconducting pairing correlations in ${\mathrm{PbTaSe}}_{2}$},\
  }\href {\doibase 10.1103/PhysRevB.93.245130} {\bibfield  {journal} {\bibinfo
  {journal} {Phys. Rev. B}\ }\textbf {\bibinfo {volume} {93}},\ \bibinfo
  {pages} {245130} (\bibinfo {year} {2016})}\BibitemShut {NoStop}%
\bibitem [{\citenamefont {Bian}\ \emph {et~al.}(2016)\citenamefont {Bian},
  \citenamefont {Chang}, \citenamefont {Sankar}, \citenamefont {Xu},
  \citenamefont {Zheng}, \citenamefont {Neupert}, \citenamefont {Chiu},
  \citenamefont {Huang}, \citenamefont {Chang}, \citenamefont {Belopolski}
  \emph {et~al.}}]{bian2016topological}%
  \BibitemOpen
  \bibfield  {author} {\bibinfo {author} {\bibfnamefont {G.}~\bibnamefont
  {Bian}}, \bibinfo {author} {\bibfnamefont {T.-R.}\ \bibnamefont {Chang}},
  \bibinfo {author} {\bibfnamefont {R.}~\bibnamefont {Sankar}}, \bibinfo
  {author} {\bibfnamefont {S.-Y.}\ \bibnamefont {Xu}}, \bibinfo {author}
  {\bibfnamefont {H.}~\bibnamefont {Zheng}}, \bibinfo {author} {\bibfnamefont
  {T.}~\bibnamefont {Neupert}}, \bibinfo {author} {\bibfnamefont {C.-K.}\
  \bibnamefont {Chiu}}, \bibinfo {author} {\bibfnamefont {S.-M.}\ \bibnamefont
  {Huang}}, \bibinfo {author} {\bibfnamefont {G.}~\bibnamefont {Chang}},
  \bibinfo {author} {\bibfnamefont {I.}~\bibnamefont {Belopolski}},  \emph
  {et~al.},\ }\bibfield  {title} {\bibinfo {title} {Topological nodal-line
  fermions in spin-orbit metal ${\mathrm{PbTaSe}}_{2}$},\ }\href@noop {}
  {\bibfield  {journal} {\bibinfo  {journal} {Nat. Commun.}\ }\textbf {\bibinfo
  {volume} {7}},\ \bibinfo {pages} {1} (\bibinfo {year} {2016})}\BibitemShut
  {NoStop}%
\bibitem [{\citenamefont {Fu}\ and\ \citenamefont {Kane}(2008)}]{fu08}%
  \BibitemOpen
  \bibfield  {author} {\bibinfo {author} {\bibfnamefont {L.}~\bibnamefont
  {Fu}}\ and\ \bibinfo {author} {\bibfnamefont {C.~L.}\ \bibnamefont {Kane}},\
  }\bibfield  {title} {\bibinfo {title} {Superconducting Proximity Effect and
  Majorana Fermions at the Surface of a Topological Insulator},\ }\href
  {\doibase 10.1103/PhysRevLett.100.096407} {\bibfield  {journal} {\bibinfo
  {journal} {Phys. Rev. Lett.}\ }\textbf {\bibinfo {volume} {100}},\ \bibinfo
  {eid} {096407} (\bibinfo {year} {2008})}\BibitemShut {NoStop}%
\bibitem [{\citenamefont {Chen}\ \emph {et~al.}(2016)\citenamefont {Chen},
  \citenamefont {Chang},\ and\ \citenamefont {Jeng}}]{chen16a}%
  \BibitemOpen
  \bibfield  {author} {\bibinfo {author} {\bibfnamefont {P.-J.}\ \bibnamefont
  {Chen}}, \bibinfo {author} {\bibfnamefont {T.-R.}\ \bibnamefont {Chang}}, \
  and\ \bibinfo {author} {\bibfnamefont {H.-T.}\ \bibnamefont {Jeng}},\
  }\bibfield  {title} {\bibinfo {title} {Ab initio study of the
  ${\mathrm{PbTaSe}}_{2}$-related superconducting topological metals},\ }\href
  {\doibase 10.1103/PhysRevB.94.165148} {\bibfield  {journal} {\bibinfo
  {journal} {Phys. Rev. B}\ }\textbf {\bibinfo {volume} {94}},\ \bibinfo
  {pages} {165148} (\bibinfo {year} {2016})}\BibitemShut {NoStop}%
\bibitem [{\citenamefont {Taylor}\ \emph {et~al.}(2007)\citenamefont {Taylor},
  \citenamefont {Carrington},\ and\ \citenamefont {Schlueter}}]{taylo07}%
  \BibitemOpen
  \bibfield  {author} {\bibinfo {author} {\bibfnamefont {O.~J.}\ \bibnamefont
  {Taylor}}, \bibinfo {author} {\bibfnamefont {A.}~\bibnamefont {Carrington}},
  \ and\ \bibinfo {author} {\bibfnamefont {J.~A.}\ \bibnamefont {Schlueter}},\
  }\bibfield  {title} {\bibinfo {title} {Specific-Heat Measurements of the Gap
  Structure of the Organic Superconductors
  $\ensuremath{\kappa}\mathrm{\text{\ensuremath{-}}}(\mathrm{ET}{)}_{2}\mathrm{Cu}[\mathrm{N}(\mathrm{CN}{)}_{2}]\mathrm{Br}$
  and
  $\ensuremath{\kappa}\mathrm{\text{\ensuremath{-}}}(\mathrm{ET}{)}_{2}\mathrm{Cu}(\mathrm{NCS}{)}_{2}$},\
  }\href {\doibase 10.1103/PhysRevLett.99.057001} {\bibfield  {journal}
  {\bibinfo  {journal} {Phys. Rev. Lett.}\ }\textbf {\bibinfo {volume} {99}},\
  \bibinfo {pages} {057001} (\bibinfo {year} {2007})}\BibitemShut {NoStop}%
\bibitem [{\citenamefont {Kresse}\ and\ \citenamefont
  {Furthm{\"u}ller}(1996)}]{RN429}%
  \BibitemOpen
  \bibfield  {author} {\bibinfo {author} {\bibfnamefont {G.}~\bibnamefont
  {Kresse}}\ and\ \bibinfo {author} {\bibfnamefont {J.}~\bibnamefont
  {Furthm{\"u}ller}},\ }\bibfield  {title} {\bibinfo {title} {Efficient
  iterative schemes for ab initio total-energy calculations using a plane-wave
  basis set},\ }\href {\doibase 10.1103/PhysRevB.54.11169} {\bibfield
  {journal} {\bibinfo  {journal} {Phys. Rev. B}\ }\textbf {\bibinfo {volume}
  {54}},\ \bibinfo {pages} {11169} (\bibinfo {year} {1996})}\BibitemShut
  {NoStop}%
\bibitem [{\citenamefont {Kresse}\ and\ \citenamefont {Hafner}(1993)}]{RN425}%
  \BibitemOpen
  \bibfield  {author} {\bibinfo {author} {\bibfnamefont {G.}~\bibnamefont
  {Kresse}}\ and\ \bibinfo {author} {\bibfnamefont {J.}~\bibnamefont
  {Hafner}},\ }\bibfield  {title} {\bibinfo {title} {Ab initio molecular
  dynamics for liquid metals},\ }\href {\doibase 10.1103/PhysRevB.47.558}
  {\bibfield  {journal} {\bibinfo  {journal} {Phys. Rev. B}\ }\textbf {\bibinfo
  {volume} {47}},\ \bibinfo {pages} {558} (\bibinfo {year} {1993})}\BibitemShut
  {NoStop}%
\bibitem [{\citenamefont {Bl{\"o}chl}(1994)}]{RN424}%
  \BibitemOpen
  \bibfield  {author} {\bibinfo {author} {\bibfnamefont {P.~E.}\ \bibnamefont
  {Bl{\"o}chl}},\ }\bibfield  {title} {\bibinfo {title} {Projector
  augmented-wave method},\ }\href {\doibase 10.1103/PhysRevB.50.17953}
  {\bibfield  {journal} {\bibinfo  {journal} {Phys. Rev. B}\ }\textbf {\bibinfo
  {volume} {50}},\ \bibinfo {pages} {17953} (\bibinfo {year}
  {1994})}\BibitemShut {NoStop}%
\bibitem [{\citenamefont {Kresse}\ and\ \citenamefont {Joubert}(1999)}]{RN449}%
  \BibitemOpen
  \bibfield  {author} {\bibinfo {author} {\bibfnamefont {G.}~\bibnamefont
  {Kresse}}\ and\ \bibinfo {author} {\bibfnamefont {D.}~\bibnamefont
  {Joubert}},\ }\bibfield  {title} {\bibinfo {title} {From ultrasoft
  pseudopotentials to the projector augmented-wave method},\ }\href {\doibase
  10.1103/PhysRevB.59.1758} {\bibfield  {journal} {\bibinfo  {journal} {Phys.
  Rev. B}\ }\textbf {\bibinfo {volume} {59}},\ \bibinfo {pages} {1758}
  (\bibinfo {year} {1999})}\BibitemShut {NoStop}%
\bibitem [{\citenamefont {Perdew}\ \emph {et~al.}(1996)\citenamefont {Perdew},
  \citenamefont {Burke},\ and\ \citenamefont {Ernzerhof}}]{RN453}%
  \BibitemOpen
  \bibfield  {author} {\bibinfo {author} {\bibfnamefont {J.~P.}\ \bibnamefont
  {Perdew}}, \bibinfo {author} {\bibfnamefont {K.}~\bibnamefont {Burke}}, \
  and\ \bibinfo {author} {\bibfnamefont {M.}~\bibnamefont {Ernzerhof}},\
  }\bibfield  {title} {\bibinfo {title} {Generalized Gradient Approximation
  Made Simple},\ }\href {\doibase 10.1103/PhysRevLett.77.3865} {\bibfield
  {journal} {\bibinfo  {journal} {Phys. Rev. Lett.}\ }\textbf {\bibinfo
  {volume} {77}},\ \bibinfo {pages} {3865} (\bibinfo {year}
  {1996})}\BibitemShut {NoStop}%
\bibitem [{\citenamefont {Perdew}\ \emph {et~al.}(1997)\citenamefont {Perdew},
  \citenamefont {Burke},\ and\ \citenamefont {Ernzerhof}}]{RN435}%
  \BibitemOpen
  \bibfield  {author} {\bibinfo {author} {\bibfnamefont {J.~P.}\ \bibnamefont
  {Perdew}}, \bibinfo {author} {\bibfnamefont {K.}~\bibnamefont {Burke}}, \
  and\ \bibinfo {author} {\bibfnamefont {M.}~\bibnamefont {Ernzerhof}},\
  }\bibfield  {title} {\bibinfo {title} {Erratum: Generalized Gradient
  Approximation Made Simple},\ }\href {\doibase 10.1103/PhysRevLett.78.1396}
  {\bibfield  {journal} {\bibinfo  {journal} {Phys. Rev. Lett.}\ }\textbf
  {\bibinfo {volume} {78}},\ \bibinfo {pages} {1396} (\bibinfo {year}
  {1997})}\BibitemShut {NoStop}%
\bibitem [{\citenamefont {Ganose}\ \emph {et~al.}(2021)\citenamefont {Ganose},
  \citenamefont {Searle}, \citenamefont {Jain},\ and\ \citenamefont
  {Griffin}}]{RN23109}%
  \BibitemOpen
  \bibfield  {author} {\bibinfo {author} {\bibfnamefont {A.~M.}\ \bibnamefont
  {Ganose}}, \bibinfo {author} {\bibfnamefont {A.}~\bibnamefont {Searle}},
  \bibinfo {author} {\bibfnamefont {A.}~\bibnamefont {Jain}}, \ and\ \bibinfo
  {author} {\bibfnamefont {S.~M.}\ \bibnamefont {Griffin}},\ }\bibfield
  {title} {\bibinfo {title} {IFermi: A python library for Fermi surface
  generation and analysis},\ }\href {\doibase 10.21105/joss.03089} {\bibfield
  {journal} {\bibinfo  {journal} {Journal of Open Source Software}\ }\textbf
  {\bibinfo {volume} {6}},\ \bibinfo {pages} {3089} (\bibinfo {year}
  {2021})}\BibitemShut {NoStop}%
\bibitem [{\citenamefont {Zheng}()}]{RN23110}%
  \BibitemOpen
  \bibfield  {author} {\bibinfo {author} {\bibfnamefont {Q.}~\bibnamefont
  {Zheng}},\ }\href {https://github.com/QijingZheng/VaspBandUnfolding}
  {\bibinfo {title} {VaspBandUnfolding},\ }\BibitemShut {NoStop}%
\bibitem [{\citenamefont {Yang}\ \emph {et~al.}(2018)\citenamefont {Yang},
  \citenamefont {Wang}, \citenamefont {Li}, \citenamefont {Bai}, \citenamefont
  {Ma}, \citenamefont {Sun}, \citenamefont {Tao}, \citenamefont {Dong},\ and\
  \citenamefont {Xu}}]{yang18c}%
  \BibitemOpen
  \bibfield  {author} {\bibinfo {author} {\bibfnamefont {X.}~\bibnamefont
  {Yang}}, \bibinfo {author} {\bibfnamefont {M.}~\bibnamefont {Wang}}, \bibinfo
  {author} {\bibfnamefont {Y.}~\bibnamefont {Li}}, \bibinfo {author}
  {\bibfnamefont {H.}~\bibnamefont {Bai}}, \bibinfo {author} {\bibfnamefont
  {J.}~\bibnamefont {Ma}}, \bibinfo {author} {\bibfnamefont {X.}~\bibnamefont
  {Sun}}, \bibinfo {author} {\bibfnamefont {Q.}~\bibnamefont {Tao}}, \bibinfo
  {author} {\bibfnamefont {C.}~\bibnamefont {Dong}}, \ and\ \bibinfo {author}
  {\bibfnamefont {Z.-A.}\ \bibnamefont {Xu}},\ }\bibfield  {title} {\bibinfo
  {title} {Superconductivity in a misfit compound
  (PbSe)$_{1.12}$(TaSe$_{2}$)},\ }\href {\doibase 10.1088/1361-6668/aae7b6}
  {\bibfield  {journal} {\bibinfo  {journal} {Superconductor Science and
  Technology}\ }\textbf {\bibinfo {volume} {31}},\ \bibinfo {pages} {125010}
  (\bibinfo {year} {2018})}\BibitemShut {NoStop}%
\bibitem [{\citenamefont {Sankar}\ \emph {et~al.}(2017)\citenamefont {Sankar},
  \citenamefont {Rao}, \citenamefont {Muthuselvam}, \citenamefont {Chang},
  \citenamefont {Jeng}, \citenamefont {Murugan}, \citenamefont {Lee},\ and\
  \citenamefont {Chou}}]{sankar2017anisotropic}%
  \BibitemOpen
  \bibfield  {author} {\bibinfo {author} {\bibfnamefont {R.}~\bibnamefont
  {Sankar}}, \bibinfo {author} {\bibfnamefont {G.~N.}\ \bibnamefont {Rao}},
  \bibinfo {author} {\bibfnamefont {I.~P.}\ \bibnamefont {Muthuselvam}},
  \bibinfo {author} {\bibfnamefont {T.-R.}\ \bibnamefont {Chang}}, \bibinfo
  {author} {\bibfnamefont {H.}~\bibnamefont {Jeng}}, \bibinfo {author}
  {\bibfnamefont {G.~S.}\ \bibnamefont {Murugan}}, \bibinfo {author}
  {\bibfnamefont {W.-L.}\ \bibnamefont {Lee}}, \ and\ \bibinfo {author}
  {\bibfnamefont {F.}~\bibnamefont {Chou}},\ }\bibfield  {title} {\bibinfo
  {title} {Anisotropic superconducting property studies of single crystal
  ${\mathrm{PbTaSe}}_{2}$},\ }\href@noop {} {\bibfield  {journal} {\bibinfo
  {journal} {J. Phys. Condens. Matter.}\ }\textbf {\bibinfo {volume} {29}},\
  \bibinfo {pages} {095601} (\bibinfo {year} {2017})}\BibitemShut {NoStop}%
\bibitem [{\citenamefont {Ziman}(2001)}]{ziman01}%
  \BibitemOpen
  \bibfield  {author} {\bibinfo {author} {\bibfnamefont {J.~M.}\ \bibnamefont
  {Ziman}},\ }\href@noop {} {\bibinfo {title} {Electrons and Phonons}}\
  (\bibinfo  {publisher} {Oxford University Press},\ \bibinfo {year}
  {2001})\BibitemShut {NoStop}%
\bibitem [{\citenamefont {Long}\ \emph {et~al.}(2016)\citenamefont {Long},
  \citenamefont {Zhao}, \citenamefont {Wang}, \citenamefont {Yang},
  \citenamefont {Li}, \citenamefont {Zi}, \citenamefont {Ren}, \citenamefont
  {Ren},\ and\ \citenamefont {Chen}}]{long2016single}%
  \BibitemOpen
  \bibfield  {author} {\bibinfo {author} {\bibfnamefont {Y.-J.}\ \bibnamefont
  {Long}}, \bibinfo {author} {\bibfnamefont {L.-X.}\ \bibnamefont {Zhao}},
  \bibinfo {author} {\bibfnamefont {P.-P.}\ \bibnamefont {Wang}}, \bibinfo
  {author} {\bibfnamefont {H.-X.}\ \bibnamefont {Yang}}, \bibinfo {author}
  {\bibfnamefont {J.-Q.}\ \bibnamefont {Li}}, \bibinfo {author} {\bibfnamefont
  {H.}~\bibnamefont {Zi}}, \bibinfo {author} {\bibfnamefont {Z.-A.}\
  \bibnamefont {Ren}}, \bibinfo {author} {\bibfnamefont {C.}~\bibnamefont
  {Ren}}, \ and\ \bibinfo {author} {\bibfnamefont {G.-F.}\ \bibnamefont
  {Chen}},\ }\bibfield  {title} {\bibinfo {title} {Single crystal growth and
  physical property characterization of non-centrosymmetric superconductor
  ${\mathrm{PbTaSe}}_{2}$},\ }\href@noop {} {\bibfield  {journal} {\bibinfo
  {journal} {Chin. Phys. Lett.}\ }\textbf {\bibinfo {volume} {33}},\ \bibinfo
  {pages} {037401} (\bibinfo {year} {2016})}\BibitemShut {NoStop}%
\bibitem [{\citenamefont {Baeva}\ \emph {et~al.}(2024)\citenamefont {Baeva},
  \citenamefont {Kolbatova}, \citenamefont {Titova}, \citenamefont {Saha},
  \citenamefont {Boltasseva}, \citenamefont {Bogdanov}, \citenamefont
  {Shalaev}, \citenamefont {Semenov}, \citenamefont {Goltsman},\ and\
  \citenamefont {Khrapai}}]{baeva24}%
  \BibitemOpen
  \bibfield  {author} {\bibinfo {author} {\bibfnamefont {E.}~\bibnamefont
  {Baeva}}, \bibinfo {author} {\bibfnamefont {A.}~\bibnamefont {Kolbatova}},
  \bibinfo {author} {\bibfnamefont {N.}~\bibnamefont {Titova}}, \bibinfo
  {author} {\bibfnamefont {S.}~\bibnamefont {Saha}}, \bibinfo {author}
  {\bibfnamefont {A.}~\bibnamefont {Boltasseva}}, \bibinfo {author}
  {\bibfnamefont {S.}~\bibnamefont {Bogdanov}}, \bibinfo {author}
  {\bibfnamefont {V.~M.}\ \bibnamefont {Shalaev}}, \bibinfo {author}
  {\bibfnamefont {A.}~\bibnamefont {Semenov}}, \bibinfo {author} {\bibfnamefont
  {G.~N.}\ \bibnamefont {Goltsman}}, \ and\ \bibinfo {author} {\bibfnamefont
  {V.}~\bibnamefont {Khrapai}},\ }\bibfield  {title} {\bibinfo {title} {Natural
  width of the superconducting transition in epitaxial TiN films},\ }\href
  {\doibase 10.1088/1361-6668/ad74a1} {\bibfield  {journal} {\bibinfo
  {journal} {Superconductor Science and Technology}\ }\textbf {\bibinfo
  {volume} {37}},\ \bibinfo {pages} {105017} (\bibinfo {year}
  {2024})}\BibitemShut {NoStop}%
\bibitem [{\citenamefont {Sun}\ \emph {et~al.}(2020)\citenamefont {Sun},
  \citenamefont {Kittaka}, \citenamefont {Sakakibara}, \citenamefont {Machida},
  \citenamefont {Sankar}, \citenamefont {Xu}, \citenamefont {Zhou},
  \citenamefont {Xing}, \citenamefont {Shi}, \citenamefont {Pyon} \emph
  {et~al.}}]{sun2020fully}%
  \BibitemOpen
  \bibfield  {author} {\bibinfo {author} {\bibfnamefont {Y.}~\bibnamefont
  {Sun}}, \bibinfo {author} {\bibfnamefont {S.}~\bibnamefont {Kittaka}},
  \bibinfo {author} {\bibfnamefont {T.}~\bibnamefont {Sakakibara}}, \bibinfo
  {author} {\bibfnamefont {K.}~\bibnamefont {Machida}}, \bibinfo {author}
  {\bibfnamefont {R.}~\bibnamefont {Sankar}}, \bibinfo {author} {\bibfnamefont
  {X.}~\bibnamefont {Xu}}, \bibinfo {author} {\bibfnamefont {N.}~\bibnamefont
  {Zhou}}, \bibinfo {author} {\bibfnamefont {X.}~\bibnamefont {Xing}}, \bibinfo
  {author} {\bibfnamefont {Z.}~\bibnamefont {Shi}}, \bibinfo {author}
  {\bibfnamefont {S.}~\bibnamefont {Pyon}},  \emph {et~al.},\ }\bibfield
  {title} {\bibinfo {title} {Fully gapped superconductivity without sign
  reversal in the topological superconductor ${\mathrm{PbTaSe}}_{2}$},\
  }\href@noop {} {\bibfield  {journal} {\bibinfo  {journal} {Phys. Rev. B}\
  }\textbf {\bibinfo {volume} {102}},\ \bibinfo {pages} {024517} (\bibinfo
  {year} {2020})}\BibitemShut {NoStop}%
\bibitem [{\citenamefont {Bouquet}\ \emph {et~al.}(2001)\citenamefont
  {Bouquet}, \citenamefont {Wang}, \citenamefont {Fisher}, \citenamefont
  {Hinks}, \citenamefont {Jorgensen}, \citenamefont {Junod},\ and\
  \citenamefont {Phillips}}]{bouquet2001phenomenological}%
  \BibitemOpen
  \bibfield  {author} {\bibinfo {author} {\bibfnamefont {F.}~\bibnamefont
  {Bouquet}}, \bibinfo {author} {\bibfnamefont {Y.}~\bibnamefont {Wang}},
  \bibinfo {author} {\bibfnamefont {R.}~\bibnamefont {Fisher}}, \bibinfo
  {author} {\bibfnamefont {D.}~\bibnamefont {Hinks}}, \bibinfo {author}
  {\bibfnamefont {J.}~\bibnamefont {Jorgensen}}, \bibinfo {author}
  {\bibfnamefont {A.}~\bibnamefont {Junod}}, \ and\ \bibinfo {author}
  {\bibfnamefont {N.}~\bibnamefont {Phillips}},\ }\bibfield  {title} {\bibinfo
  {title} {Phenomenological two-gap model for the specific heat of
  ${\mathrm{MgB}}_{2}$},\ }\href@noop {} {\bibfield  {journal} {\bibinfo
  {journal} {Europhys. Lett.}\ }\textbf {\bibinfo {volume} {56}},\ \bibinfo
  {pages} {856} (\bibinfo {year} {2001})}\BibitemShut {NoStop}%
\bibitem [{\citenamefont {Carrington}\ and\ \citenamefont
  {Manzano}(2003)}]{carrington2003magnetic}%
  \BibitemOpen
  \bibfield  {author} {\bibinfo {author} {\bibfnamefont {A.}~\bibnamefont
  {Carrington}}\ and\ \bibinfo {author} {\bibfnamefont {F.}~\bibnamefont
  {Manzano}},\ }\bibfield  {title} {\bibinfo {title} {Magnetic penetration
  depth of ${\mathrm{MgB}}_{2}$},\ }\href@noop {} {\bibfield  {journal}
  {\bibinfo  {journal} {Physica C}\ }\textbf {\bibinfo {volume} {385}},\
  \bibinfo {pages} {205} (\bibinfo {year} {2003})}\BibitemShut {NoStop}%
\bibitem [{\citenamefont {Khasanov}\ \emph {et~al.}(2008)\citenamefont
  {Khasanov}, \citenamefont {Shengelaya}, \citenamefont {Maisuradze},
  \citenamefont {Di~Castro}, \citenamefont {Savi{\'c}}, \citenamefont
  {Weyeneth}, \citenamefont {Park}, \citenamefont {Jang}, \citenamefont {Lee},\
  and\ \citenamefont {Keller}}]{khasanov2008nodeless}%
  \BibitemOpen
  \bibfield  {author} {\bibinfo {author} {\bibfnamefont {R.}~\bibnamefont
  {Khasanov}}, \bibinfo {author} {\bibfnamefont {A.}~\bibnamefont
  {Shengelaya}}, \bibinfo {author} {\bibfnamefont {A.}~\bibnamefont
  {Maisuradze}}, \bibinfo {author} {\bibfnamefont {D.}~\bibnamefont
  {Di~Castro}}, \bibinfo {author} {\bibfnamefont {I.}~\bibnamefont
  {Savi{\'c}}}, \bibinfo {author} {\bibfnamefont {S.}~\bibnamefont {Weyeneth}},
  \bibinfo {author} {\bibfnamefont {M.}~\bibnamefont {Park}}, \bibinfo {author}
  {\bibfnamefont {D.}~\bibnamefont {Jang}}, \bibinfo {author} {\bibfnamefont
  {S.-I.}\ \bibnamefont {Lee}}, \ and\ \bibinfo {author} {\bibfnamefont
  {H.}~\bibnamefont {Keller}},\ }\bibfield  {title} {\bibinfo {title} {Nodeless
  superconductivity in the infinite-layer electron-doped cuprate superconductor
  Sr$_{0.9}$La$_{0.1}$CuO$_{2}$},\ }\href@noop {} {\bibfield  {journal}
  {\bibinfo  {journal} {Phys. Rev. B}\ }\textbf {\bibinfo {volume} {77}},\
  \bibinfo {pages} {184512} (\bibinfo {year} {2008})}\BibitemShut {NoStop}%
\bibitem [{\citenamefont {McMillan}(1968)}]{mcmil68}%
  \BibitemOpen
  \bibfield  {author} {\bibinfo {author} {\bibfnamefont {W.~L.}\ \bibnamefont
  {McMillan}},\ }\bibfield  {title} {\bibinfo {title} {Transition Temperature
  of Strong-Coupled Superconductors},\ }\href@noop {} {\bibfield  {journal}
  {\bibinfo  {journal} {Phys. Rev.}\ }\textbf {\bibinfo {volume} {167}},\
  \bibinfo {pages} {331} (\bibinfo {year} {1968})}\BibitemShut {NoStop}%
\bibitem [{\citenamefont {Bouvier}\ \emph {et~al.}(1991)\citenamefont
  {Bouvier}, \citenamefont {Lethuillier},\ and\ \citenamefont
  {Schmitt}}]{bouvi91}%
  \BibitemOpen
  \bibfield  {author} {\bibinfo {author} {\bibfnamefont {M.}~\bibnamefont
  {Bouvier}}, \bibinfo {author} {\bibfnamefont {P.}~\bibnamefont
  {Lethuillier}}, \ and\ \bibinfo {author} {\bibfnamefont {D.}~\bibnamefont
  {Schmitt}},\ }\bibfield  {title} {\bibinfo {title} {Specific heat in some
  gadolinium compounds. I. Experimental},\ }\href {\doibase
  10.1103/PhysRevB.43.13137} {\bibfield  {journal} {\bibinfo  {journal} {Phys.
  Rev. B}\ }\textbf {\bibinfo {volume} {43}},\ \bibinfo {pages} {13137}
  (\bibinfo {year} {1991})}\BibitemShut {NoStop}%
\bibitem [{\citenamefont {Henheik}\ \emph {et~al.}()\citenamefont {Henheik},
  \citenamefont {Langmann},\ and\ \citenamefont {Lauritsen}}]{henhe24}%
  \BibitemOpen
  \bibfield  {author} {\bibinfo {author} {\bibfnamefont {J.}~\bibnamefont
  {Henheik}}, \bibinfo {author} {\bibfnamefont {E.}~\bibnamefont {Langmann}}, \
  and\ \bibinfo {author} {\bibfnamefont {A.~B.}\ \bibnamefont {Lauritsen}},\
  }\bibfield  {title} {\bibinfo {title} {Multi-band superconductors have
  enhanced critical temperatures},\ }\href@noop {} {\bibinfo  {journal}
  {arXiv:2409.17297}\ }\BibitemShut {NoStop}%
\bibitem [{\citenamefont {Yu}\ \emph {et~al.}(2024)\citenamefont {Yu},
  \citenamefont {Ciccarino}, \citenamefont {Bianco}, \citenamefont {Errea},
  \citenamefont {Narang},\ and\ \citenamefont {Bernevig}}]{yu24}%
  \BibitemOpen
\bibfield  {journal} {  }\bibfield  {author} {\bibinfo {author} {\bibfnamefont
  {J.}~\bibnamefont {Yu}}, \bibinfo {author} {\bibfnamefont {C.~J.}\
  \bibnamefont {Ciccarino}}, \bibinfo {author} {\bibfnamefont {R.}~\bibnamefont
  {Bianco}}, \bibinfo {author} {\bibfnamefont {I.}~\bibnamefont {Errea}},
  \bibinfo {author} {\bibfnamefont {P.}~\bibnamefont {Narang}}, \ and\ \bibinfo
  {author} {\bibfnamefont {B.~A.}\ \bibnamefont {Bernevig}},\ }\bibfield
  {title} {\bibinfo {title} {Non-trivial quantum geometry and the strength of
  electron--phonon coupling},\ }\href@noop {} {\bibfield  {journal} {\bibinfo
  {journal} {Nat. Phys.}\ }\textbf {\bibinfo {volume} {20}},\ \bibinfo {pages}
{1262} (\bibinfo {year} {2024})}\BibitemShut {NoStop}%
\bibitem{da} https://doi.org/10.5281/zenodo.16047123.
\end{thebibliography}
\end{document}